\pdfoutput=1

\documentclass[twoside,leqno,twocolumn]{article}

\usepackage[letterpaper]{geometry}
\usepackage{siamproceedings}

\usepackage{graphicx} 
\usepackage{amsopn,amsfonts}
\usepackage{hyperref}
\hypersetup{
  pdftitle={Expand \& Reduce: a Framework Bridging Branch-and-Reduce and Dynamic Programming},
  pdfauthor={Martin Vatshelle, Kenneth Langedal, Ernestine Grossmann},
  pdfsubject={Maximum Weight Independent Set},
  pdfkeywords={Maximum Weight Independent Set, branch-and-reduce, dynamic programming, data reduction, edge expansion}
}
\usepackage{xspace}
\usepackage[table]{xcolor}
\usepackage{url}
\usepackage{adjustbox}
\usepackage{pgfplots}
\usepackage{booktabs}
\usepackage{pdflscape}
\usepackage{tabularx}
\usepackage{array}
\usepackage[linesnumbered,ruled,vlined,algo2e]{algorithm2e}
\pgfplotsset{compat=1.18}
\usepackage{nameref}
\usepackage{rotating}
\usepackage{ragged2e}
\newcolumntype{L}[1]{>{\RaggedRight\arraybackslash}m{#1}}

\newtheorem{regel}{Rule}

\newsiamremark{remark}{Remark}
\newsiamthm{claim}{Claim}

\newcommand{\thistheoremname}{???}
\newtheorem*{genericthm*}{\thistheoremname}
\newcounter{genericthm}

\makeatletter
\newenvironment{namedthm*}[2][] 
  {\def\currentthmname{#2}%
   \protected@edef\@currentlabelname{\currentthmname}%
   \refstepcounter{genericthm}%
   \if\relax\detokenize{#1}\relax
     \begin{genericthm*}[#2]
   \else
     \begin{genericthm*}[#2 (#1)]
   \fi}
  {\end{genericthm*}}
\makeatother
  
\newcommand{\I}{I}
\newcommand{\enr}{expand-and-reduce}
\newcommand{\etal}{et~al.}
\newcommand{\w}{\omega}
\newcommand*\mis[1]{{\alpha(#1)}}
\def\indSet{{\sc Maximum Weighted Independent Set\xspace}}
\usepackage{tikz}
\usepackage{xcolor}
\usetikzlibrary{ arrows.meta, positioning, tikzmark, calc}
\usepackage{pgfplots}
\usepgfplotslibrary{colorbrewer}
\pgfdeclarelayer{foreground}
\pgfsetlayers{main,foreground} 

\tikzset{
	graph/.style={draw, fill=remainingGColor, rounded corners=3mm, inner sep=1.6mm, align=center, minimum width=2.5cm},
	node/.style={circle, draw, fill=black, inner sep=0pt, minimum size=6pt},
	lnode/.style={node, label=left:#1},
	rnode/.style={node, label=right:#1},
	tnode/.style={node, label=above:#1},
	bnode/.style={node, label=below:#1},
	nodeR/.style={ node, draw=lightgray, fill=lightgray},
	nodeE/.style={ node, draw=redgray, fill=redgray, fill opacity=0.6, draw opacity=0.6},
	nodeI/.style={ node, draw=greengray, fill=greengray, fill opacity=0.6, draw opacity=0.6},
	weight/.style={node, rectangle, fill opacity=0, draw opacity=0, text opacity = 1,text=weightColor, yshift=0.3cm, font=\small},
	uweight/.style={node, rectangle, fill opacity=0, draw opacity=0, text opacity = 1,text=weightColor, yshift=-0.3cm, font=\small},
	lweight/.style={node, rectangle, fill opacity=0, draw opacity=0, text opacity = 1,text=weightColor, xshift=-0.3cm, font=\small},
	rweight/.style={node, rectangle, fill opacity=0, draw opacity=0, text opacity = 1,text=weightColor, xshift=0.35cm, font=\small},
	weightR/.style={node, rectangle, fill opacity=0, draw opacity=0, text opacity = 1,text=redgray, yshift=0.5cm, font=\small},
	weightText/.style={rectangle, fill opacity=1, text opacity = 1,text=weightColor, font=\small},
	edge/.style={draw=black, thick},
	edgeR/.style={ edge, draw=lightgray, dashed} 
}

\definecolor{remainingGColor}{RGB}{226,243,253}
\definecolor{weightColor}{RGB}{204,0,0}
\definecolor{darkgreen}{rgb}{0.0, 0.5, 0.0}
\definecolor{lightgreen}{rgb}{0.235, 0.7, 0.235}
\definecolor{green}{RGB}{0, 150, 130}
\definecolor{green70}{RGB}{76, 181, 167}
\definecolor{blue}{RGB}{70, 100, 170}
\definecolor{blue70}{RGB}{125,146,195}
\definecolor{blue50}{RGB}{162,177,212}                                          \definecolor{blue30}{RGB}{199,208,229}
\definecolor{blue15}{RGB}{227,232,242}
\definecolor{lightgray}{rgb}{0.86,0.86,0.86}
\definecolor{greengray}{RGB}{193,228,224}
\definecolor{redgray}{RGB}{233,183,183}
\definecolor{red}{RGB}{204,0,0}

\newcommand{\remainingG}[2]{
	\begin{pgfonlayer}{foreground}
		\node[graph] at (#1) (box) {#2};
	\end{pgfonlayer}
}


\begin{document}

\title{\Large Expand \& Reduce: a Framework Bridging Branch-and-Reduce and Dynamic Programming\\
\vspace{1ex}\textup{\large Applied to the Maximum Weight Independent Set Problem}}

\newcommand{\orcid}[1]{\href{https://orcid.org/#1}{\includegraphics[width=8pt]{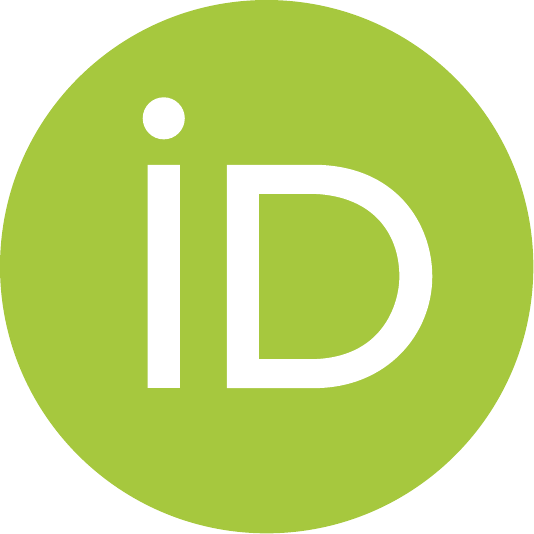}}}

\author{
  Martin Vatshelle\,\orcid{0009-0009-1788-2509}\thanks{\textbf{Corresponding author}, \email{Martin.Vatshelle@uib.no}, University of Bergen, Norway.}
  \and
  Kenneth Langedal\,\orcid{0009-0001-6838-4640}\thanks{Heidelberg University, Germany.}
  \and
  Ernestine Gro{\ss}mann\,\orcid{0000-0002-9678-0253}\footnotemark[2]
}

\date{}

\maketitle

\pagenumbering{arabic}
\setcounter{page}{1}
\fancyfoot[C]{\small\thepage}

\begin{abstract}
We introduce \emph{expand-and-reduce}, a framework that augments classical reduction rules with a new type of optimality-preserving operation, called an \emph{expansion rule}. 
While a reduction rule in practice typically makes the instance smaller, an expansion rule may enlarge the instance yet still making progress towards a solution.
We present a simple expansion rule called \emph{edge expansion} for the \textsc{Maximum Weighted Independent Set} (MWIS) problem. It encodes the joint choice of two non-adjacent vertices by adding an edge between them and creating a new vertex representing their union. 
We prove that the number of edge expansions needed to transform a graph into a clique equals the number of independent sets in the graph minus the number of vertices.
We demonstrate that expand-and-reduce, using edge expansion and just two well-known reduction rules, has greater reduction power than a set of 22 commonly used reduction rules for MWIS. 
Therefore, it can significantly simplify methods that depend on reduction rules. To highlight the practical potential of the expand-and-reduce framework further, we provide an implementation and preliminary results showing that a simple implementation outperforms a far more advanced method for MWIS preprocessing.
Moreover, expand-and-reduce can simulate both branching and dynamic-programming-style state construction, yielding exact and fixed-parameter algorithms with matching asymptotic running times. By that, expand-and-reduce offers a unified and conceptually clean perspective on techniques that are usually considered incompatible.
\end{abstract}


\vfill
\noindent{
\begin{minipage}{\dimexpr\columnwidth-2\fboxsep-2\fboxrule\relax}
\footnotesize\RaggedRight
\textbf{Artifact \& Supplementary Material}\\[0.4ex]
$\triangleright$~\textbf{Code:} \href{https://github.com/KennethLangedal/ExpandAndReduce}{\nolinkurl{github.com/KennethLangedal/ExpandAndReduce}}\\
$\triangleright$~\textbf{Archived at:} \url{https://doi.org/10.5281/zenodo.22960898}\\
$\triangleright$~\textbf{Dataset:} \url{https://doi.org/10.5281/zenodo.22661369}
\end{minipage}%
}

\vfill
\newpage

\section{Introduction.}

For $\mathsf{NP}$-hard graph problems, two of the most successful algorithmic paradigms are \emph{branch-and-reduce} and \emph{dynamic programming}.
Both are powerful, but they organize computation in very different ways.
Branch-and-reduce works directly on the given instance and repeatedly applies reductions, resorting to branching only when no further reduction is possible, pruning branches not leading to optimality.
Dynamic programming, in contrast, typically introduces a larger family of partial solutions, utilizing structure to compare and eliminate partial solutions not leading to optimality.
Because of this difference, the two paradigms are often developed separately, even when they address essentially the same problem.

In this paper, we introduce a framework that sits naturally between branch-and-reduce and dynamic programming, which we call \emph{\enr}.
The basic idea is to augment classical reduction rules with a new type of optimality-preserving operation, called an \emph{expansion rule}.
Whereas a reduction rule simplifies an instance by making it smaller, an expansion rule may enlarge the instance in order to expose structure that trigger reduction rules.
Exhaustively applying reduction rules leads to an irreducible instance, while expansion rules can be applied until a solvable instance is reached.
Thus, expansion rules play a role analogous to the introduction of partial solutions in dynamic programming, while staying within the same problem class enables the reduction-based progress of branch-and-reduce.
In particular, when reduction rules are exhausted, expansion rules can still be applied to make progress, and could replace branching and dynamic programming altogether.
Expansion and reduction rules can be applied interchangeably throughout the algorithm, allowing the expanded instance to trigger further reductions in the same way as in branch-and-reduce.

We develop this idea for the \textsc{Maximum Weight Independent Set} (MWIS) problem by introducing a new expansion rule we call \emph{edge expansion}. Given two non-adjacent vertices $a$ and $b$, edge expansion creates one new vertex $v_{ab}$ with neighbors $N[a] \cup N[b]$ and adds the edge $\{a,b\}$, this new vertex is representing the union of $a$ and $b$ and is given the weight $\w(a) + \w(b)$.
We prove that exhaustively applying this operation to an $n$-vertex graph $G$ transforms $G$ into a clique. We also show in this paper that each vertex can be labeled with a different independent set of $G$, so no matter which edge expansions we do, the number of edge expansions is at most the number of distinct independent sets in $G$, minus $n$.
Since there are easy reduction rules that can reduce vertices in a clique, we will always reach a single vertex whose weight is the maximum weight independent set in the graph.

Our work is inspired by Gellner et al.~\cite{gellner2021boosting} who recently introduced increasing transformations as a step towards a branch-and-transform paradigm for MWIS. The first increasing transformations were introduced by Ebenegger~\etal~\cite{ebenegger1984pseudo} for the unweighted problem. The concept of temporarily increasing an instance during preprocessing was also explored by Figiel et al.~\cite{figiel2022there} for Vertex Cover by ``undoing'' previously applied data reductions to escape local reduction minima.

Our aim is not merely to define a new graph operation, but to show that it provides a simple language for understanding several existing algorithmic ideas.
When combining edge expansion with two well-known reduction rules, we get a framework for solving the MWIS problem.
In order to create an algorithm we only need to determine the order of reductions and pairs of vertices to apply edge expansion on.
We demonstrate the reduction power of this simple technique by proving that a sequence of these three rules can cover commonly used reduction rules described in a recent survey by Gro{\ss}mann et al.~\cite{grossmann2024comprehensivesurveydatareduction}.
This shows that these extra reduction rules do not provide additional reduction power beyond our three rules, thereby partially addressing Open Problem 1 by Abu-Khzam et al.~\cite{Abu-Khzam2022}.

When reduction rules are applied exhaustively, we get an irreducible instance. To make progress on this input, the most common technique, both in practice and in theory, is to apply branching.
We show that for many branch-and-reduce algorithms solving MWIS in $O^*(c^n)$ time, we can also solve MWIS in $O^*(c^n)$ time using edge expansions and reduction rules and hence avoid branching altogether. We achieve this by doing edge expansions in an order easily derived from the branching algorithm.

Fixed-parameter tractable (FPT) algorithms define a parameter of structure in a graph that assigns a value to each graph.
For practical purposes, it is beneficial to pair the right parameter with the right problem so that easily solvable instances are assigned low values and hard-to-solve instances are assigned high values.
Then, we can design an algorithm that is polynomial in the input size but exponential in the parameter assigned, which implies exponential worst-case running time but polynomial time for all inputs assigned a constant value.
The graph parameter boolean-width and the problem MWIS is one such good pair. One way to measure the quality of such pairs is by analyzing how large the class of inputs is that can be solved in polynomial time. See~\cite{belmonte2013graph, boolw1} for reasoning why this pair is considered good.
When designing FPT algorithms, dynamic programming is the most commonly used technique.
We prove that our algorithm can simulate the dynamic programming algorithm for boolean-width. Because boolean-width compares favorably to other parameters~\cite{vatshelle2012thesis}, we get as a corollary that we can simulate standard dynamic programming algorithms for tree-width, rank-width, clique-width, mim-width, and many other parameters.

The idea of combining branching and dynamic programming goes back more than 50 years. 
One of the earlier ideas is the notion of dominance relation \cite{helman1989combine,ibaraki1977power,kohler1974characterization}. 
This is basically the fundamental idea that one partial solution can be better than another. 
By memorizing all partial solutions during a branch-and-bound execution, we can prune some branches.
The idea of dominance relations is strikingly similar to Bellman's principle of optimality, which led to dynamic programming~\cite{bellman1954dp}.
Fomin et al.~\cite{fomin2009two}, observed that branching is best when the input has many high-degree vertices, and dynamic programming is best when all vertices are of low degree. 
Therefore, running branch-and-bound until all large-degree vertices have been deleted, and then switching to dynamic programming, gave improved results.
These results have all been on the theoretical side, and none of them have led to algorithms used in practice.

\textbf{Our Contributions.} We show the power of the \enr{} framework, in particular our algorithm technique for MWIS. 
We prove that if we can find the right order to apply the expansion rule, this technique is more powerful than both branching and dynamic programming.
Furthermore, we provide a simple yet flexible algorithmic framework that is suitable for practical applications.
We have created a reference implementation that randomly applies edge expansions followed by reduction rules, and reverts if no reduction in graph size is achieved. This is deliberately not a fast algorithm, as more than 99\% of the edge expansions are undone. Nevertheless, this algorithm still reduces further than the previous state-of-the-art preprocessing method.
Finding a better way to guide the selection of edge expansion pairs can significantly reduce execution time.

\section{Preliminaries.}

A graph $G=(V,E,\w)$ is an undirected vertex-weighted graph with ${n=|V|}$ and ${m = |E|}$, where ${V =\{0,\ldots,n-1\}}$ and $\w : V \rightarrow \mathbb{R}^{+}$. The neighborhood $N(v)$ of a vertex $v \in V$ is defined as $N(v) = \{u \in V \mid \{u,v\} \in E\}$.
Additionally, we define $N[v]=N(v) \cup \{v\}$. This is generalized for sets by the neighborhood $N(U)$ of a set of vertices ${U \subset V}$, i.e.\ ${N(U) = \cup_{v \in U} N(v)\setminus U}$ and $N[U] = N(U) \cup U$.
For a graph $G=(V,E,\w)$ we define an induced subgraph $G[S]$  on a subset of vertices $S \subset V$ by $G[S] = (S,\{\{u,v\} \in E \mid u,v\in S\},\w)$. For $S \subset V$, we use the notation $G-S$ instead of $G[V\setminus S]$. Similarly for a single vertex $v\in V$, we abbreviate $G[V\setminus\{v\}]$ to $G-v$.

A set $\I\subseteq V$ is called an \textit{independent set} (IS) if for all vertices $u,v \in \I$ there is no edge $\{u,v\} \in E$.
The \textsc{Maximum Independent Set} (MIS) problem is that of finding an IS with maximum cardinality.
Similarly, the \indSet{} (MWIS) problem is that of finding an IS of maximum weight. The weight of an IS $\I$ is defined as $\omega(\I) = \sum_{v \in \I}\omega(v)$ and $\alpha(G)$ denotes the weight of an MWIS of $G$. A \textit{clique} is a set $C \subseteq V$ such that all vertices are pairwise adjacent. A vertex is \textit{simplicial} when its neighborhood forms a clique. A vertex $v$ is removable if $\mis{G} = \mis{G - v}$, while it is includable if $\mis{G} = \omega(v) + \mis{G - N[v]}$.

\section{The Expand \& Reduce Framework for MWIS.}
\label{sec_edge_expansion}

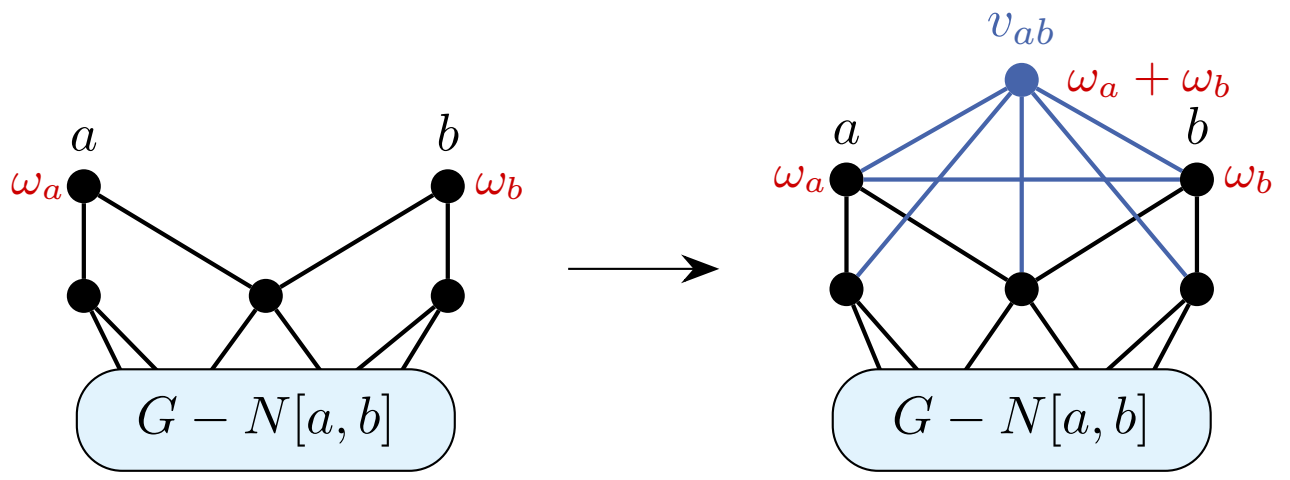
\begin{figure}
	\centering
	\begin{tikzpicture}
		\node (w) at (5,4.25) {};
		\node[tnode=$b$] (v) [below right=.5cm and 1cm of w] {};
		\node[tnode=$a$] (u) [below left=.5cm and 1cm of w] {};
		\node[node] (v2) [below =.5cm of v] {};
		\node[node] (u1) [below =.5cm of u] {};
       	\node[node] (v1) at ($(v2)!0.5!(u1)$) {};
		\remainingG{5,2}{$G-N[a,b]$}
		\draw[edge] (v) -- (v1);
		\draw[edge] (v) -- (v2);
		\draw[edge] (u) -- (v1);
		\draw[edge] (u) -- (u1);
		\draw[edge] (4.2,2) -- (u1);
		\draw[edge] (4.6,2) -- (u1);
		\draw[edge] (4.4,2) -- (v1);
		\draw[edge] (5.6,2) -- (v1);
		\draw[edge] (5.2,2) -- (v2);
		\draw[edge] (5.7,2) -- (v2);
		\node[rweight] (vw) at (v) {$\w_b$};
		\node[lweight] (uw) at (u) {$\w_a$};        
        \draw[-{Stealth[scale=1.5]}] (7.,3.0) to (8.,3.);
	    \begin{scope}[xshift=5cm]
		\node[tnode={\textcolor{blue}{$v_{ab}$}},blue] (w) at (5,4.25) {};
		\node[tnode=$b$] (v) [below right=.5cm and 1cm of w] {};
		\node[tnode=$a$] (u) [below left=.5cm and 1cm of w] {};
		\node[node] (v2) [below =.5cm of v] {};
		\node[node] (u1) [below =.5cm of u] {};
       	\node[node] (v1) at ($(v2)!0.5!(u1)$) {};
		\remainingG{5,2}{$G-N[a,b]$}
		\draw[edge] (v) -- (v1);
		\draw[edge] (v) -- (v2);
		\draw[edge] (u) -- (v1);
		\draw[edge] (u) -- (u1);
		\draw[edge] (4.2,2) -- (u1);
		\draw[edge] (4.6,2) -- (u1);
		\draw[edge] (4.4,2) -- (v1);
		\draw[edge] (5.6,2) -- (v1);
		\draw[edge] (5.2,2) -- (v2);
		\draw[edge] (5.7,2) -- (v2);
		\draw[edge,blue] (v) -- (u);
		\draw[edge,blue] (v) -- (w);
		\draw[edge,blue] (u) -- (w);
		\draw[edge,blue] (v1) -- (w);
		\draw[edge,blue] (u1) -- (w);
		\draw[edge,blue] (v2) -- (w);
		\node[rweight] (vw) at (v) {$\w_b$};
		\node[lweight] (uw) at (u) {$\w_a$};   
		\node[rweight,xshift=0.5cm] (ww) at (w) {$\w_a+\w_b$};   
	    \end{scope}
        
	\end{tikzpicture}
	\caption{Illustration for edge expansion. The rule is applied to the vertices $a$ and $b$ with weights $\w_a, \w_b$ respectively. The expansion added the blue vertex and edges in the right graph.}\label{fig:edge_expansion}
\end{figure}

This section introduces the expand-and-reduce framework for MWIS. This includes the expansion rule, edge expansion, and two reduction rules. The main idea of edge expansion is to encode independent sets into the graph as new vertices, illustrated in Figure~\ref{fig:edge_expansion}. 

\begin{namedthm*}{Edge Expansion}\label{edge_expansion}
    Let $a,b\in V$ be non-adjacent vertices in $G$. Edge expansion is a transformation of $G$ denoted $G' = \text{Expand}(G,a,b)$ where $G'$ is obtained by adding the edge $\{a,b\}$ and a new vertex $v_{ab}$ connected to $N[a]\cup N[b]$ with weight $\omega(a)+\omega(b)$. 
\end{namedthm*}

\begin{theorem}[Edge Expansion Preserves Independent Sets]\label{thm:edge_expansion}
    Let $G=(V,E,\w)$ be a graph, $a,b \in V$ be non-adjacent vertices, and $G' = (V',E',\w') = \text{Expand}(G,a,b)$, then there is a one-to-one mapping between independent sets in $G$ and $G'$, preserving weight. It holds that $\mis{G} = \mis{G'}$.
\end{theorem}
\begin{proof}
    Any independent set that does not contain both $a$ and $b$ is an independent set in both $G$ and $G'$.
    For an independent set $I \subseteq V$ where $a,b \in I$ we construct $I' = (I \setminus \{a,b\}) \cup \{v_{ab}\}$. 
    Since $(I \setminus \{a,b\}) \cap (N[a]\cup N[b]) = \emptyset$ and $N[v_{ab}] = N[a]\cup N[b]$ we know $I'$ is an independent set.
    For an independent set $I' \subseteq V'$ we know that $I'$ can not contain both $a$ and $b$ since edge expansion added an edge between them so the only time $I'$ does not exist in $G$ is if $v_{ab} \in I'$.
    We construct $I = (I' \setminus \{v_{ab}\}) \cup \{a,b\}$.
    Since $v_{ab}$ is adjacent to $N[a]\cup N[b]$ we know that $I$ is an independent set.
    In each case we either replace $v_{ab}$ with $\{a,b\}$ or replace $\{a,b\}$ with $v_{ab}$, and since $\w(v_{ab}) = \w(a) + \w(b)$ weights are preserved for all independent sets.
\end{proof}


\begin{remark}
    Each original vertex $v \in V(G)$ can be viewed as representing the singleton independent set $\{v\}$. Inductively, when edge expansion is performed on two non-adjacent vertices representing disjoint independent sets $I_a$ and $I_b$, the new vertex $v_{ab}$ represents the independent set $I_a \cup I_b$ with weight $\w(v_{ab}) = \w(I_a \cup I_b) = \w(a) + \w(b)$.
\end{remark}

While the graph $G$ is not a clique, we can find non-adjacent vertices and perform an edge expansion.
A natural question to ask is if we can continue doing edge expansions indefinitely, or if not, how many edge expansions can we do before $G$ turns into a clique?
We show that any set of vertices $S\subseteq V$ can always be transformed into a clique by edge expansions.

\begin{lemma}[Clique Expansion]\label{lem:clique_expansion}
    Let $G = (V,E,\w)$ be a graph, $S \subseteq V$, and initialize $S' = S$ and $G' = G$. While there exist two non-adjacent vertices $a,b \in S'$, set $G'= \text{Expand}(G',a,b)$, and $S'=S' \cup \{v_{ab}\}$. If no further such pair exists, $G'[S']$ is a clique where for every independent set $I \subseteq S$, $S'$ contains one vertex $v_I$ with $\w(v_I) = \w(I)$ and $N[v_I] = S' \cup \bigcup_{v \in I}N[v]$.
\end{lemma}
\begin{proof}
    By Theorem \ref{thm:edge_expansion}, there is a one-to-one mapping between independent sets before and after an edge expansion, by applying this multiple times we see that there is a one-to-one mapping between independent sets in $G$ and $G'$.
    Since the mapping only changes the vertices involved in an edge expansion, there must also be a one-to-one mapping between independent sets in $G[S]$ and $G'[S']$ and this mapping preserves the neighbors outside $S$, i.e. $\forall I \subseteq S, I' \subseteq S'$ such that $I$ is mapped to $I'$ we have $N(I) \cap \overline{S} = N(I') \cap \overline{S'}$.
    Let $M$ be the number of independent sets in $G[S]$, if there exist some independent set in $G[S]$ that map to an independent set of size at least $2$ we could have applied more edge expansions.
    Therefore, we must at some point reach $M = |S'|$ and because there is a one-to-one mapping we conclude $G'[S']$ is a clique of size $M$. 
\end{proof}

Next we describe the reduction rules that are part of our framework for MWIS. These are commonly used in solvers for the MWIS problem~\cite{grossmann2024comprehensivesurveydatareduction}. 
They were introduced for the weighted independent set problem in the branch-and-reduce solver \textsc{KaMIS} by Lamm~et~al.~\cite{lamm2019exactly}. 
The reduction rule domination describes a pattern, where a vertex $u$ is always to be preferred over one of its neighbors $v$.
In this case, we say $u$ dominates $v$.

\begin{namedthm*}[by Lamm~\etal~\cite{lamm2019exactly}]{Domination}\label{domination}
    Let $u,v\in V$ be adjacent in $G$, such that $N[u]\subseteq N[v]$ and $\omega(u)\geq \omega(v)$. 
    Then the vertex $v$ is removable, and we obtain $G' = G - v$.
\end{namedthm*}

In addition to includable and removable vertices, some reduction rules postpone decisions or combine vertices into new vertices. We call this procedure \textit{folding}. Note that here, we need to \textit{unfold} the solution obtained on the reduced graph $G'$ to obtain a solution for the original graph $G$. More specifically, a folding reduction returns a solution \textit{offset} value such that $\mis{G} = \textit{offset} + \mis{G'}$, along with a function that maps a solution from $G'$ back to $G$.

The second rule we want to discuss was originally introduced by Lamm~\etal~\cite{lamm2019exactly} under the name \textit{Simplicial Weight Transfer} and uses the concept of folding. We call it \nameref{clique_fold} instead, since we think this better describes the rule. It considers simplicial vertices, that is, vertices $v\in V$ where their neighborhood $N(v)$ forms a clique. For any simplicial vertex $v$, there is exactly one vertex from $N[v]$ in an MWIS. 

\begin{namedthm*}[by Lamm~\etal~\cite{lamm2019exactly}]{Clique Fold}\label{clique_fold}
    Let $v\in V$ be simplicial. 
    Any vertex $u \in N(v)$ such that $\w(u) \leq \w(v)$ can be removed by domination. Therefore, we assume this is done first. Afterwards, we know $v$ is the vertex in $N[v]$ with the lowest weight and we can fold $v$ into its neighborhood. This is done by reducing the weight of all neighbors $u\in N(v)$ to $\omega(u) = \omega(u) - \omega(v)$. The transformed graph is $G'=G-v$. It holds that $\alpha(G) = \omega(v) + \alpha(G')$. If a solution $I'$ on $G'$ does not contain any neighbor of $v$, i.e. $I'\cap N(v) = \emptyset$, then a solution $I$ to the original graph $G$ is constructed by $I = I'\cup \{v\}$, otherwise $I = I'$.
\end{namedthm*}

\textbf{An Expand \& Reduce Algorithm for MWIS.}
To formally define an algorithm using domination, clique folding, and edge expansion, an ordering of edge expansions is needed. When applying reduction and expansion rules exhaustively, regardless of the order, these operations reduce the graph to a single vertex, thereby solving the instance exactly. Although the specific ordering does not affect the exactness of the algorithm, it significantly influences its running time. It is important to note that solving a graph through edge expansions can take an exponential number of steps. Therefore, identifying effective orderings for expansions presents a key algorithmic challenge. In the remainder of this paper, we demonstrate how reductions, fixed-parameter tractable (FPT) algorithms, and branching algorithms can all provide expansion orderings that correspond to their respective running times.

\section{Simplification and Extension of Known Reduction Rules for MWIS.}
\label{sec_reduction_rules}

\begin{table*}[t]
    \caption{List of commonly used general reductions and their special cases. For a detailed description, see~\cite{grossmann2024comprehensivesurveydatareduction}. We briefly describe each rule and which of edge expansion (EE), domination (D), and clique fold (CF) are needed to reduce the same pattern. The rules in the last row are only covered conceptually.}\label{tab:reduction_overview}
    \small
    \setlength{\extrarowheight}{1pt}
    \begin{tabular}{L{3.3cm} L{5.7cm} m{2.0cm} L{4.5cm}}
      \textbf{Rule} & \textbf{Brief Description} & \textbf{By} & \textbf{Special Cases}\\
      \midrule
   \rowcolor{gray!15}
      \textit{Simplicial Weight \newline Transfer} & \nameref{clique_fold}  & CF  & \textit{Degree 1}, \textit{Triangle}, \textit{Simplicial \newline Vertex} \\
      \textit{Heavy Vertex} & Rule~\ref{regel:heavy_vertex}: Include a vertex $v\in V$ with $\omega(v) \geq \alpha(G[N(v)])$. & EE + D + CF & \textit{Neighborhood Removal}, \textit{Clique Neighborhood Removal}\\
    \rowcolor{gray!15}
      \textit{Weighted Funnel} & Rule~\ref{regel:funnel}: Fold $v$ if there exists $u \in N(v)$ such that $N(v)\setminus \{u\}$ is a clique and $\omega(v) \geq \max\{\omega(x) \mid x \in N(v) \setminus \{u\}\}$. & EE + CF&\\
      \textit{Basic Single-Edge} &Rule~\ref{regel:se}: Remove $v$ if some neighbor $u\in N(v)$ has $\omega(N(u)\setminus N(v)) \leq \omega(u)$.&EE + D + CF& \textit{Domination}\\
    \rowcolor{gray!15}
      \textit{Extended Single-Edge} &Rule~\ref{regel:ese}: For $\{u,v\}\in E$, remove $N(u)\cap N(v)$ if $\w(v) \geq \w(N(v)\setminus \{u\})$.&EE + D + CF&\\
      \textit{Extended Twin} & Rule~\ref{regel:e_twin}: Include or fold two non-adjacent vertices with equal neighborhoods. & EE + D + CF & \textit{Twin}\\
    \rowcolor{gray!15}
      \textit{Struction} & Rule~\ref{regel:struction}: Remove $N[v]$ and add  a vertex for each independent set in $G[N(v)]$ heavier than $\w(v)$, made into a clique connected to neighbors outside $N[v]$. & EE + D + CF &\textit{Neighborhood Folding}, \textit{Generalized Neighborhood Folding}\\     
      \textit{Heavy Set }&Rule~\ref{regel:heavy_set}: Generalizes Rule~\ref{regel:heavy_vertex} to a set of vertices. & EE + CF&\textit{Two Vertex Neighbor Removal}, \textit{Almost Twin}, \textit{Critical Weight} \newline \textit{Independent Set}\\ 
    \rowcolor{gray!15}
      \textit{Vertex Cut} &Rule~\ref{regel:cut}: Reduces small components created by cutting articulation points.& EE + D + CF&\\ 
      \textit{Degree 2 V-Shape}& Rule~\ref{regel:Vshape}: Include or fold a vertex with only two non-adjacent neighbors.& EE + CF &\\\midrule \midrule
    \rowcolor{gray!15}
      \textit{Extended Unconfined}&Rule~\ref{regel:e_unconfined}: Remove $v$ if assuming it is in every MWIS yields a contradiction.& EE + D & \textit{Unconfined}\\
    \end{tabular}
\end{table*}

There have been many reduction rules introduced for MWIS~\cite{butenko2007using,gellner2021boosting,grossmann2025accelerating,lamm2019exactly,liu2023application,xiao2024maximum,xiao2021efficient,zheng2020efficient}. The survey by Gro{\ss}mann et al.~\cite{grossmann2024comprehensivesurveydatareduction} presents an overview of these rules. In this section, we show that these existing reductions can be simplified by using only edge expansions, dominations, and clique folds, so that the reduced instance is often identical or very similar. This can greatly simplify practical preprocessing methods and indicates that further reducible patterns can be found that are not covered by the currently known rule set. It can also simplify correctness proofs, as any rule expressed in terms of edge expansions, dominations, and clique folds is automatically correct.

For some reduction rules, there is a degree of freedom in how the reduced instance is defined, particularly with respect to edges and vertex weights. While many existing rules are primarily designed to minimize the number of vertices and edges, they therefore often remove more edges than our framework does. We argue, however, that minimizing the number of edges is not necessarily an appropriate objective, as some edges may, in fact, simplify the resulting instance.
To illustrate this, consider a graph together with the additional information that two non-adjacent vertices can never appear simultaneously in an independent set. In this case, adding an edge between these vertices is safe, as it preserves the correctness. Moreover, the added edge encodes this obtained information directly in the graph structure, thereby eliminating the need to consider independent sets containing both vertices as candidate solutions.
This observation motivates our decision not to minimize the number of edges in the reduced instance artificially. Instead, we consider two instances $G_1$ and $G_2$ \emph{equivalent} if there is a one-to-one mapping between $V(G_1)$ and $V(G_2)$ and $\alpha(G_1)=\alpha(G_2)$, even if the vertex weights and number of edges are different.

\begin{theorem}\label{thm:reductions_cover}
    \nameref{edge_expansion}, \nameref{domination}, and \nameref{clique_fold} can simplify 22 commonly used reduction rules (see Table~\ref{tab:reduction_overview}), i.e., there is a way to apply the three rules to reduce the same pattern, yielding reduced instances that are equivalent, as defined above.
\end{theorem}

Many reducible patterns are restricted to a local region of the graph, often the neighborhood of a vertex. The idea behind many reduction rules can be described by the following: either a few small independent sets in that local area dominate all other possible independent sets in that region, or, conversely, every independent set containing a specific vertex is dominated by the remaining feasible choices.
Therefore, we can often follow the same procedure to show the equivalence between existing rules and our framework. We can perform edge expansions to make the set of vertices containing the local area into a clique following Lemma~\ref{lem:clique_expansion}. With this, we add all these possible combinations of independent sets as vertices to the graph. Then, we can check for domination on these to remove suboptimal solutions. After this procedure, simplicial vertices can be removed by clique fold.

In Table~\ref{tab:reduction_overview}, we present a list of general reduction rules, together with a short description and their special cases. To prove Theorem~\ref{thm:reductions_cover}, it is sufficient to consider those general rules, since the special cases follow from that. 
Here, we present two reductions and explain in detail how they are covered by edge expansion, domination, and clique fold. For the remaining reduction rules listed in Table~\ref{tab:reduction_overview}, the proofs are given in Appendix~\ref{sec_app_finishing_proof}. 
In addition to the 22 rules equivalently covered by our framework, two more rules are covered conceptually, even though we cannot reach an exact equivalent reduced instance. Those rules are listed in the bottom part of Table~\ref{tab:reduction_overview} and discussed in Appendix~\ref{sec_app_finishing_proof}.

The first rule we examine is the heavy vertex rule introduced by Xiao~\etal~\cite{xiao2021efficient}. Here, a vertex can be included if its weight is larger than or equal to the weight of an MWIS in its neighborhood. 
This pattern is very general and covers several rules, see Table~\ref{tab:reduction_overview}.

\begin{figure*}
	\centering
	\begin{tikzpicture}
		\node[rnode=$v$] (v) at (5,4.25) {};
		\node[node] (v3) [below right=.75cm and 1cm of v] {};
		\node[node] (v1) [below left=.75cm and 1cm of v] {};
    	\node[node, yshift=-0.35cm] (v2) at ($(v1)!0.5!(v3)$) {};
		\remainingG{5,2.2}{$G-N[v]$}
		\draw[edge] (v3) -- (v);
		\draw[edge] (v2) -- (v);
		\draw[edge] (v1) -- (v);
		\draw[edge] (v2) -- (v3);
		\draw[edge] (4.2,2.2) -- (v1);
		\draw[edge] (4.6,2.2) -- (v1);
		\draw[edge] (5.2,2.2) -- (v3);
		\draw[edge] (5.7,2.2) -- (v3);
		\node[weight] (vweight) at (v) {$12$};
		\node[lweight] (v1weight) at (v1) {$4$};
		\node[uweight] (v2weight) at (v2) {$5$};
		\node[rweight] (v3weight) at (v3) {$6$};
        
        \draw[bend left=10, -{Stealth[scale=1.5]}] (5.2,4.7) to  node[midway, above] {\nameref{edge_expansion}} (8.8,4.7);
	    \begin{scope}[xshift=4cm]
    		\node[node] (v) at (5,4.25) {};
    		\node[node] (v3) [below right=.75cm and 1cm of v] {};
    		\node[node] (v1) [below left=.75cm and 1cm of v] {};
    		\node[node, yshift=-0.35cm] (v2) at ($(v1)!0.5!(v3)$) {};
    		\node[node,blue] (e1) [below left=.015cm and .75cm of v] {};
    		\node[node,green] (e2) [below right=.015cm and .75cm of v] {};
    		\remainingG{5,2.2}{$G-N[v]$}
    		\draw[edge] (v3) -- (v);
    		\draw[edge] (v2) -- (v);
    		\draw[edge] (v1) -- (v);
    		\draw[edge] (v2) -- (v3);
    		\draw[edge] (4.2,2.2) -- (v1);
    		\draw[edge] (4.6,2.2) -- (v1);
    		\draw[edge] (5.2,2.2) -- (v3);
    		\draw[edge] (5.7,2.2) -- (v3);
            \draw[blue] (e1) -- (v);
            \draw[blue] (e1) -- (v1);
            \draw[blue] (e1) -- (v2);
            \draw[blue] (e1) -- (v3);
            \draw[green] (e2) -- (v);
            \draw[green] (e2) -- (v1);
            \draw[green] (e2) -- (v2);
            \draw[green] (e2) -- (v3);
            \draw[edge,blue] (v1) -- (v2);
            \draw[edge,green] (v1) -- (v3);
            \draw[blue] (e1) -- (e2);
            
            \draw[blue] (e1) -- (4.1,2.2);
            \draw[blue] (e1) -- (4.4,2.2);
            \draw[green] (e2) -- (3.74,2.2);
            \draw[green] (e2) -- (4.0,2.2);
            \draw[green] (e2) -- (5.4,2.2);
            \draw[green] (e2) -- (5.8,2.2);
    		\node[weight] (vweight) at (v) {$12$};
    		\node[lweight] (v1weight) at (v1) {$4$};
    		\node[uweight] (v2weight) at (v2) {$5$};
    		\node[rweight] (v3weight) at (v3) {$6$};
    		\node[weight] (e1weight) at (e1) {$9$};
    		\node[weight] (e2weight) at (e2) {$10$};
        \draw[bend left=10, -{Stealth[scale=1.5]}] (5.2,4.7) to  node[midway, above] {\nameref{domination}} (8.8,4.7);
	    \end{scope}
        \begin{scope}[xshift=8cm]        
    		\node[node] (v) at (5,4.25) {};
    		\node[nodeE] (v3) [below right=.75cm and 1cm of v] {};
    		\node[nodeE] (v1) [below left=.75cm and 1cm of v] {};
    		\node[nodeE, yshift=-0.35cm] (v2) at ($(v1)!0.5!(v3)$) {};
    		\node[nodeE] (e1) [below left=.015cm and .75cm of v] {};
    		\node[nodeE] (e2) [below right=.015cm and .75cm of v] {};
    		\remainingG{5,2.2}{$G-N[v]$}
    		\draw[edgeR] (v3) -- (v);
    		\draw[edgeR] (v2) -- (v);
    		\draw[edgeR] (v1) -- (v);
    		\draw[edgeR] (v2) -- (v3);
    		\draw[edgeR] (4.2,2.2) -- (v1);
    		\draw[edgeR] (4.6,2.2) -- (v1);
    		\draw[edgeR] (5.2,2.2) -- (v3);
    		\draw[edgeR] (5.7,2.2) -- (v3);
            \draw[edgeR] (e1) -- (v);
            \draw[edgeR] (e2) -- (v);
            \draw[edgeR] (e1) -- (v1);
            \draw[edgeR] (e2) -- (v1);
            \draw[edgeR] (e1) -- (v2);
            \draw[edgeR] (e2) -- (v2);
            \draw[edgeR] (e1) -- (v3);
            \draw[edgeR] (e2) -- (v3);
            \draw[edgeR] (v1) -- (v2);
            \draw[edgeR] (v1) -- (v3);
            \draw[edgeR] (e1) -- (e2);
            
            \draw[edgeR] (e1) -- (4.1,2.2);
            \draw[edgeR] (e1) -- (4.4,2.2);
            \draw[edgeR] (e2) -- (3.74,2.2);
            \draw[edgeR] (e2) -- (4.0,2.2);
            \draw[edgeR] (e2) -- (5.4,2.2);
            \draw[edgeR] (e2) -- (5.8,2.2);
        \draw[bend left=10, -{Stealth[scale=1.5]}] (5.2,4.7) to  node[midway, above] {\nameref{clique_fold}} (8.8,4.7);
	    \end{scope}
        
	    \begin{scope}[xshift=12cm]
    		\node[nodeI] (v) at (5,4.25) {};
    		\node[nodeE] (v3) [below right=.75cm and 1cm of v] {};
    		\node[nodeE] (v1) [below left=.75cm and 1cm of v] {};
    		\node[nodeE, yshift=-0.35cm] (v2) at ($(v1)!0.5!(v3)$) {};
    		\node[nodeE] (e1) [below left=.015cm and .75cm of v] {};
    		\node[nodeE] (e2) [below right=.015cm and .75cm of v] {};
    		\remainingG{5,2.2}{$G-N[v]$}
    		\draw[edgeR] (v3) -- (v);
    		\draw[edgeR] (v2) -- (v);
    		\draw[edgeR] (v1) -- (v);
    		\draw[edgeR] (v2) -- (v3);
    		\draw[edgeR] (4.2,2.2) -- (v1);
    		\draw[edgeR] (4.6,2.2) -- (v1);
    		\draw[edgeR] (5.2,2.2) -- (v3);
    		\draw[edgeR] (5.7,2.2) -- (v3);
            \draw[edgeR] (e1) -- (v);
            \draw[edgeR] (e2) -- (v);
            \draw[edgeR] (e1) -- (v1);
            \draw[edgeR] (e2) -- (v1);
            \draw[edgeR] (e1) -- (v2);
            \draw[edgeR] (e2) -- (v2);
            \draw[edgeR] (e1) -- (v3);
            \draw[edgeR] (e2) -- (v3);
            \draw[edgeR] (v1) -- (v2);
            \draw[edgeR] (v1) -- (v3);
            \draw[edgeR] (e1) -- (e2);
            
            \draw[edgeR] (e1) -- (4.1,2.2);
            \draw[edgeR] (e1) -- (4.4,2.2);
            \draw[edgeR] (e2) -- (3.74,2.2);
            \draw[edgeR] (e2) -- (4.0,2.2);
            \draw[edgeR] (e2) -- (5.4,2.2);
            \draw[edgeR] (e2) -- (5.8,2.2);
	    \end{scope}
	\end{tikzpicture}
	\caption{Illustration of Rule~\ref{regel:heavy_vertex} (Heavy Vertex) in the expand-and-reduce framework.}\label{fig:heavy_vertex}
\end{figure*}
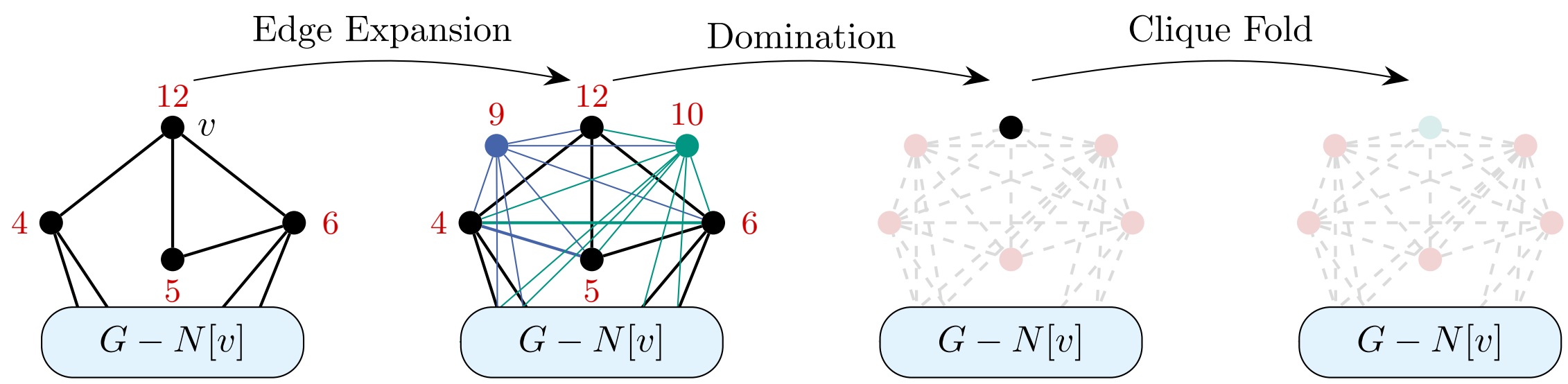

\begin{regel}[Heavy Vertex]\label{regel:heavy_vertex}
    A vertex $v\in V$ with $\omega(v) \geq \alpha(G[N(v)])$ is includable.
\end{regel}
\begin{proof}
    Let $v\in V$ with $\omega(v) \geq \alpha(G[N(v)])$. Using Lemma~\ref{lem:clique_expansion}, we can make $N(v)$ into a clique, where the highest weight vertex of this clique has weight $\alpha(G[N(v)])$. After this step, we know $v$ is simplicial. Since $\omega(v) \geq \alpha(G[N(v)])$, $v$ dominates all its neighbors. Using clique fold, $v$ can be folded and later included. An illustration of these steps is given in Figure~\ref{fig:heavy_vertex}.
\end{proof}

The second rule we want to examine applies to almost simplicial vertices. This rule relies on case distinction and considers different sets and weight adjustments. By that, it is harder to implement and to prove correctness. This is another example of the simplification that can be achieved with our new framework.

\begin{regel}[Funnel by Gro{\ss}mann et al.~\cite{grossmann2025accelerating}]\label{regel:funnel}
    Let $u,v\in V$ be adjacent vertices such that $N(v)\setminus \{u\}$ forms a clique.
    If $\omega(v) \geq \max\{\omega(x) \mid x \in N(v) \setminus \{u\}\}$, then we can reduce this pattern. 
    For that, we first define the remaining neighbors as $N'(v) = \{x \in N(v) \setminus N[u] \mid \omega(x) + \omega (u) > \omega(v)\}$. In both following cases, the offset is increased by $\omega(v)$.
    \begin{itemize}
        \item If $\omega(v) \geq \omega(u)$, fold $u$ and $v$ into their neighborhood. This results in $G' = G - (N[v]\setminus N'(v))$ and for all $x \in N'(v)$, we set $N(x) = N(x) \cup N(u)$ and $\omega(x) = \omega(x) + \omega(u) - \omega(v)$.
        \item If $\omega(v) < \omega(u)$, fold $v$ into its neighborhood. This results in $G' = G - (N[v]\setminus (N'(v)\cup \{u\}))$ with $\omega(u) = \omega(u) - \omega(v)$ and for all $x \in N'(v)$, we set $N(x) = N(x) \cup N(u)$.
    \end{itemize}
    
\end{regel}
\begin{proof}
    We start by making the neighborhood of $v$ into a clique following Lemma~\ref{lem:clique_expansion}. For that, we perform edge expansions between $u$ and vertices in $N(v)\setminus N[u]$.  
    Now, $v$ is simplicial and therefore dominates all vertices in $N(v)$ with lower or equal weight. After removing these vertices, we apply clique fold.
    In the first case, where $\omega(v) \geq \omega(u)$, all original vertices in $N[v]$ have been removed. Only the vertices that were added by edge expansion remain. These describe sets containing $u$ and one other vertex $x \in N'(v)$ and have a neighborhood of $N(x) \cup N(u)$. Their weight is $\omega(u) + \omega(x) -\omega(v)$.
    The only difference in the second case, where $\omega(v) < \omega(u)$, is that now $u$ is not dominated by $v$ and therefore remains in the reduced instance. In our framework, we still have edges between $u$ and vertices in $N'(x)$, while the original Rule~\ref{regel:funnel} does not have them. Both of the resulting reduced instances are still equivalent, according to our definition for Theorem~\ref{thm:reductions_cover}.
\end{proof}

\begin{remark}
    In Theorem~\ref{thm:reductions_cover}, we do not consider the computational cost of checking the rules. From a practical point of view, it can still make sense to use special-case rules that can be checked more efficiently.
\end{remark}

We showed that many reduction rules rely on domination relationships between different independent sets in a small local part of the graph. Previously, for each rule, there was a specific reducible pattern, with associated weight relations that had to be described in detail. This results in a long list of rules to be checked. Our framework, on the other hand, directly captures the reasoning behind the rule itself. This makes the high-level idea of the reduction rule explicit and shows that our framework covers all rules that can be derived from such domination arguments. Moreover, it opens up new possibilities for mining previously undiscovered special-case reduction rules. By this, we not only simplify existing reduction rule pipelines for the MWIS problem, but can reduce further than was possible before.

\section{Relation to FPT and Dynamic Programming.}

Fixed-parameter tractable (FPT) algorithms are defined by a measure function that assigns a value to each graph.
This measure is meant to indicate how easy or hard it is to solve a class of problems on a given graph.
E.g., if the graph is a tree or a clique many problems are easily solvable.
Dynamic programming is one of the most common techniques used for developing FPT algorithms and normally involves: 
\begin{enumerate}
    \item A graph decomposition that provides a path for solving increasingly larger subproblems.
    \item Defining an equivalence relation for subproblems.
    \item At each step, store a table with one solution for each equivalence class of subproblems.
    \item Each table entry is computed from previously computed table entries.
\end{enumerate}

When two partial solutions are considered equivalent in a table it means we only need to keep one of those solutions and the table entry for that equivalence class stores the best solution.
The running time is then mainly dependent on the number of equivalence classes.

Boolean-width~\cite{boolw1} is a graph parameter very suited to measuring the complexity of MIS, we denote boolean-width of a graph $G$ as $boolw(G)$. 
Boolean-width has a low value compared to most other graph parameters i.e. lots of graph classes have boolean-width $O(\log n)$~\cite{belmonte2013graph} and an algorithm for solving MWIS with running time $boolw(G) \cdot 2^{2 \cdot boolw(G)} \cdot n^2$ is given in~\cite{boolw1}.
Boolean-width is based on neighborhood equivalence, which is an extension of twins to sets.

\begin{definition}[Neighborhood Equivalent~\cite{boolw1}]\label{def:neighborhoodEq}
Given a graph $G = (V,E,\w)$ and a cut $(A,\overline{A})$
Let $S_1, S_2 \subseteq A$ be two subsets.
We define $S_1 \equiv_A S_2$ if $N[S_1] \cap \overline{A} = N[S_2] \cap \overline{A}$.
\end{definition}


\begin{theorem}
    Let $G = (V,E,\w)$ be a graph and $k = boolw(G)$ then the number of edge expansions and dominations needed to reduce $G$ to a single vertex is at most $2^{2k} \cdot n$. 
\end{theorem}
\begin{proof}
    Since the graph has boolean-width $k$, there exists a boolean-decomposition $(T,\delta)$ of width $k$ where $T$ is a binary tree and $\delta$ is a one-to-one mapping between leaves of $T$ and $V(G)$. 
    We arbitrarily root $T$ and perform clique expansions (see Lemma~\ref{lem:clique_expansion}) bottom up from the leaves towards the root. 
    Recall that each vertex in $G$ created by edge expansions represents a subset of the original vertices in $G$.
    For leaves, $S(v)$ is defined to be the vertex mapped by $\delta$, for each internal vertex $v \in T$ with children $a$ and $b$, we define the subset $S(v) = S(a) \cup S(b)$. 
    We assume that both $S(a)$ and $S(b)$ are cliques, by doing clique expansion on $S(v)$, we can turn $S(v)$ into a clique of size at most $|S(a)| \cdot |S(b)|$.
    Each vertex in $S(v)$ represents an independent set, and by neighborhood equivalence, two subsets are equivalent with respect to $S(v)$ if they have the same neighborhood outside $S(v)$. 
    Since $S(v)$ is a clique, independent sets are single vertices and having the same neighborhood across the cut means these will be true twins, and one of them will be removed by domination and we continue up the tree.
    Since $(T,\delta)$ is a boolean decomposition of width $k$, both $S(a)$ and $S(b)$ will be cliques of size at most $2^k$.
    To do a clique expansion of $S(v)$ we need at most $2^{2k}$ edge expansions before we use reductions to reduce $S(v)$ back down to $2^k$ vertices. 
    There is at most $n-1$ internal nodes in $T$ so the total number of edge expansions needed is $2^{2k}(n-1)$.
\end{proof}

Although we here prove that the number of edge expansions is low as a function of boolean-width, a similar argument works for most other graph parameters, e.g., tree-width, clique-width, rank-width, mim-width~\cite{vatshelle2012thesis}, and one-sided-mim-width~\cite{Bergougnoux2023oneside}.

\section{Relation to Branching.}

Branch-and-bound and its variants—such as branch-and-reduce and branch-and-cut—are popular techniques for solving $\mathsf{NP}$-hard problems in both theory and practice~\cite{fomin2006measure, kamis2017}. 
Using branch-and-bound to solve MWIS involves selecting a branching vertex $u$ and generating two subproblems: one in which $u \in I$ and one in which $u \notin I$. This process recurses until the subproblems become empty. To avoid an exhaustive search of the solution space, pruning discards branches that cannot contain a better solution than the best seen so far. Branch-and-reduce extends branch-and-bound by applying reduction rules before each branching step, further shrinking the search tree.

To demonstrate that expand-and-reduce is as powerful as branch-and-bound, we use edge expansions to construct a graph, where each added vertex corresponds to a node in the branching tree. The procedure is described in Algorithm~\ref{alg:bnb1} and~\ref{alg:bnb2} and shown in Figure~\ref{fig:branching}.

\begin{theorem}\label{thm:branching}
    Let $A$ be a branching algorithm for the MWIS problem that can be adapted to \enr{}\footnote{Using only reduction rules we have shown to be covered by domination, clique fold, and edge expansion, upper/lower bound pruning, and branching.}, and suppose its search tree contains $T=\mathcal{O}(\alpha^{\mu(G)})$ nodes for some graph measure $\mu$. Then modifying $A$ to use edge expansions instead of branching, according to Algorithm~\ref{alg:bnb1}, produces at most $\mathcal{O}(\alpha^{\mu(G)})$ vertices representing branching states.
\end{theorem}

\begin{proof}
    We maintain a one-to-one correspondence between nodes of the search tree of $A$ and vertices created in Algorithm~\ref{alg:bnb2}. The root branch is represented by the initial vertex added in Algorithm~\ref{alg:bnb1}. Each branching step replaces a state vertex $b$ by two successor states, where one continues to be represented by $b$ and the other is represented by a newly created vertex $v_{ab}$. Thus, every branching step introduces exactly one new vertex. Since a search tree with $T$ nodes has $T-1$ branching steps (edges), the total number of new vertices is at most $T$. 
\end{proof}

\begin{algorithm2e}
\DontPrintSemicolon
\SetCommentSty{itshape}
\caption{Branch-and-bound in the \enr{} framework.}\label{alg:bnb1}
\SetKwFunction{UB}{UB}
\SetKwFunction{SELECT}{Select}
\KwIn{A graph $G=(V,E,\w)$, a branching rule $\SELECT \colon \mathcal{G} \to V$, and an upper bound function $\UB \colon \mathcal{G} \to \mathbb{N}$, where $\mathcal{G}$ is the set of all graphs}
\KwOut{Maximum weight of an independent set in $G$}
\tcp*[l]{$r$ is the root branch}
Add $r$ to $G$ with $\w(r) \gets 0$ and $N(r) \gets \emptyset$\;
\Return \textsc{Branch}$(G, r, 0)$\;
\end{algorithm2e}

\begin{algorithm2e}
\DontPrintSemicolon
\SetCommentSty{itshape}
\caption{\textsc{Branch}$(G, b, \mathrm{LB})$: the \enr{} branching function. }\label{alg:bnb2}
\KwIn{A graph $G=(V,E,\w)$, a branching vertex $b$, and the lower bound $\mathrm{LB}$}
\KwOut{Updated lower bound}

\eIf{$\mathrm{LB} \ge \w(b) + \texttt{UB}(G - N[b])$}{
\Return $\mathrm{LB}$
}{
\If{$G - N[b] = \emptyset$}{
\Return $\max(\mathrm{LB}, \w(b))$
}

$a \gets \texttt{Select}(G - N[b])$\;
$G' \gets \texttt{Expand}(G, a, b)$\;
\tcp*[l]{$v_{ab}$ is the include branch}
$\mathrm{LB} \gets \texttt{Branch}(G', v_{ab}, \mathrm{LB})$\;
\tcp*[l]{$b$ is the exclude branch}
\Return $\texttt{Branch}(G', b, \mathrm{LB})$\;

}
\end{algorithm2e}

\begin{remark}\label{rem:branching-clique}
    The set of vertices representing branching states always forms a clique. For a branching vertex $b$ the graph $G - N[b]$ is the subproblem in that branch. To save memory, this clique can be stored as a list rather than explicitly storing all its edges.
\end{remark}

\newlength{\figbranchheight}\setlength{\figbranchheight}{9.5cm}
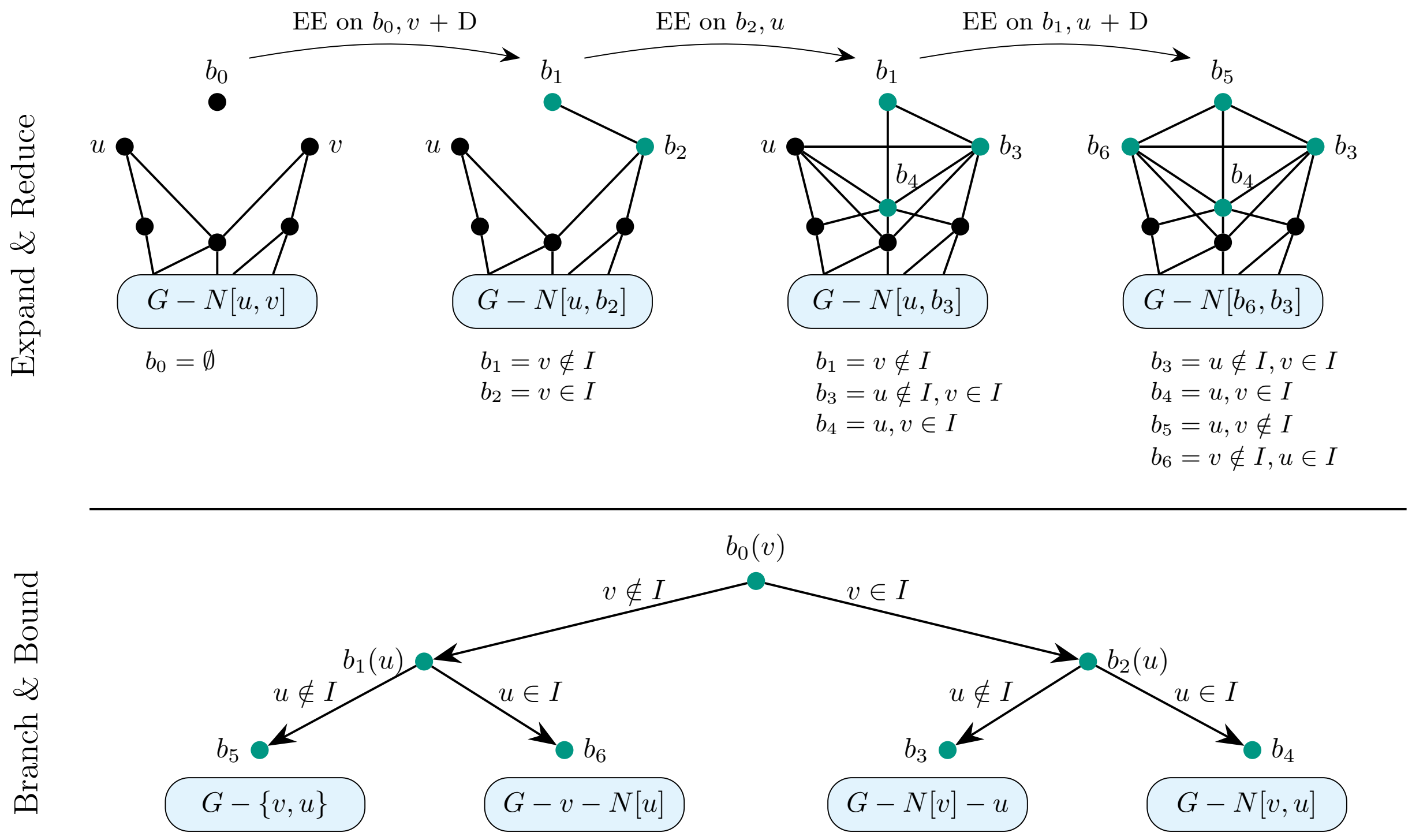
\begin{figure*}[t]
	\centering
    \begin{adjustbox}{max width=\textwidth, max totalheight=\figbranchheight}
	\begin{tikzpicture}
        \node[rotate=90] (l0) at (2.6,3.2) {\large Expand \& Reduce};

		\node[tnode=$b_0$] (b0) at (5,5) {};
		\node[lnode=$u$] (u) [below left=.4cm and 1cm of b0] {};
		\node[rnode=$v$] (v) [below right=.4cm and 1cm of b0] {};
    	\node[node] (v1) [below left=1.4cm and .75cm of b0] {};
    	\node[node] (v3) [below right=1.4cm and .75cm of b0] {};
    	\node[node] (v2) at ($(v1)!0.5!(v3) - (0,0.2)$) {};
		\remainingG{5,2.5}{$G-N[u, v]$}
		\draw[edge] (u) -- (v1);
		\draw[edge] (u) -- (v2);
		\draw[edge] (v) -- (v2);
		\draw[edge] (v) -- (v3);
        \draw[edge] (v1) -- (4.2,0 |- box.north);
        \draw[edge] (v2) -- (5.0,0 |- box.north);
        \draw[edge] (v2) -- (4.2,0 |- box.north);
        \draw[edge] (v3) -- (5.7,0 |- box.north);
        \draw[edge] (v3) -- (5.2,0 |- box.north);
        \draw[bend left=10, -{Stealth[scale=1.5]}] (5.4,5.55) to  node[midway, above] {\small EE on $b_0,v$ + D} (8.8,5.55);

        \node[text width=3.0cm] (t0) [below left=2.9cm and -2.3cm of b0] {\small ${b_0=\emptyset}$};

	    \begin{scope}[xshift=4.2cm]
        \node[tnode=$b_1$, green] (b0) at (5,5) {};
		\node[lnode=$u$] (u) [below left=.4cm and 1cm of b0] {};
		\node[rnode=$b_2$, green] (v) [below right=.4cm and 1cm of b0] {};
    	\node[node] (v1) [below left=1.4cm and .75cm of b0] {};
    	\node[node] (v3) [below right=1.4cm and .75cm of b0] {};
    	\node[node] (v2) at ($(v1)!0.5!(v3) - (0,0.2)$) {};
		\remainingG{5,2.5}{$G-N[u, b_2]$}
		\draw[edge] (v) -- (b0);
		\draw[edge] (u) -- (v1);
		\draw[edge] (u) -- (v2);
		\draw[edge] (v) -- (v2);
		\draw[edge] (v) -- (v3);
        \draw[edge] (v1) -- (4.2,0 |- box.north);
        \draw[edge] (v2) -- (5.0,0 |- box.north);
        \draw[edge] (v2) -- (4.2,0 |- box.north);
        \draw[edge] (v3) -- (5.7,0 |- box.north);
        \draw[edge] (v3) -- (5.2,0 |- box.north);
        \draw[bend left=10, -{Stealth[scale=1.5]}] (5.4,5.55) to  node[midway, above] {\small EE on $b_2,u$} (8.8,5.55);

        \node[text width=3.0cm] (t0) [below left=2.9cm and -2.3cm of b0] {\small ${b_1=v\notin I}$};
        \node[text width=3.0cm] (t1) [below left=3.3cm and -2.3cm of b0] {\small ${b_2=v\in I}$};
	    \end{scope}

	    \begin{scope}[xshift=8.4cm]
        \node[tnode=$b_1$, green] (b0) at (5,5) {};
		\node[lnode=$u$] (u) [below left=.4cm and 1cm of b0] {};
		\node[rnode=$b_3$, green] (v) [below right=.4cm and 1cm of b0] {};
		\node[node, label={[inner sep=1pt, shift={(0.25,0.1)}]above:$b_4$}, green] (w) [below=1.1cm of b0] {};
    	\node[node] (v1) [below left=1.4cm and .75cm of b0] {};
    	\node[node] (v3) [below right=1.4cm and .75cm of b0] {};
    	\node[node] (v2) at ($(v1)!0.5!(v3) - (0,0.2)$) {};
		\remainingG{5,2.5}{$G-N[u, b_3]$}
		\draw[edge] (v) -- (b0);
		\draw[edge] (u) -- (v1);
		\draw[edge] (u) -- (v2);
		\draw[edge] (v) -- (v2);
		\draw[edge] (v) -- (v3);
		\draw[edge] (v) -- (w);
		\draw[edge] (v) -- (u);
		\draw[edge] (w) -- (u);
		\draw[edge] (w) -- (b0);
		\draw[edge] (w) -- (v1);
		\draw[edge] (w) -- (v2);
		\draw[edge] (w) -- (v3);
        \draw[edge] (v1) -- (4.2,0 |- box.north);
        \draw[edge] (v2) -- (5.0,0 |- box.north);
        \draw[edge] (v2) -- (4.2,0 |- box.north);
        \draw[edge] (v3) -- (5.7,0 |- box.north);
        \draw[edge] (v3) -- (5.2,0 |- box.north);
        \draw[bend left=10, -{Stealth[scale=1.5]}] (5.4,5.55) to  node[midway, above] {\small EE on $b_1,u$ + D} (8.8,5.55);

        \node[text width=3.0cm] (t0) [below left=2.9cm and -2.3cm of b0] {\small ${b_1=v\notin I}$};
        \node[text width=3.0cm] (t1) [below left=3.3cm and -2.3cm of b0] {\small ${b_3=u\notin I, v\in I}$};
        \node[text width=3.0cm] (t2) [below left=3.7cm and -2.3cm of b0] {\small ${b_4=u,v\in I}$};
	    \end{scope}

	    \begin{scope}[xshift=12.6cm]
        \node[tnode=$b_5$, green] (b0) at (5,5) {};
		\node[lnode=$b_6$, green] (u) [below left=.4cm and 1cm of b0] {};
		\node[rnode=$b_3$, green] (v) [below right=.4cm and 1cm of b0] {};
		\node[node, label={[inner sep=1pt, shift={(0.25,0.1)}]above:$b_4$}, green] (w) [below=1.1cm of b0] {};
    	\node[node] (v1) [below left=1.4cm and .75cm of b0] {};
    	\node[node] (v3) [below right=1.4cm and .75cm of b0] {};
    	\node[node] (v2) at ($(v1)!0.5!(v3) - (0,0.2)$) {};
		\remainingG{5,2.5}{$G-N[b_6, b_3]$}
		\draw[edge] (v) -- (b0);
		\draw[edge] (u) -- (b0);
		\draw[edge] (u) -- (w);
		\draw[edge] (u) -- (v1);
		\draw[edge] (u) -- (v2);
		\draw[edge] (v) -- (v2);
		\draw[edge] (v) -- (v3);
		\draw[edge] (v) -- (w);
		\draw[edge] (v) -- (u);
		\draw[edge] (w) -- (b0);
		\draw[edge] (w) -- (v1);
		\draw[edge] (w) -- (v2);
		\draw[edge] (w) -- (v3);
        \draw[edge] (v1) -- (4.2,0 |- box.north);
        \draw[edge] (v2) -- (5.0,0 |- box.north);
        \draw[edge] (v2) -- (4.2,0 |- box.north);
        \draw[edge] (v3) -- (5.7,0 |- box.north);
        \draw[edge] (v3) -- (5.2,0 |- box.north);

        \node[text width=3.0cm] (t0) [below left=2.9cm and -2.3cm of b0]  {\small ${b_3=u\notin I, v\in I}$};
        \node[text width=3.0cm] (t1) [below left=3.3cm and -2.3cm of b0]  {\small ${b_4=u,v\in I}$};
        \node[text width=3.0cm] (t2) [below left=3.7cm and -2.3cm of b0]  {\small ${b_5=u,v\notin I}$};
        \node[text width=3.0cm] (t3) [below left=4.1cm and -2.3cm of b0]  {\small ${b_6=v\notin I, u\in I}$};
	    \end{scope}

        \draw[edge] (3.4,-0.1) -- (19.9,-0.1);

	    \begin{scope}[yshift=-3.0cm]

            \node[rotate=90] (l1) at (2.6,0.6) {\large Branch \& Bound};

            \node[tnode=${b_0}(v)$, green] (v0) at (11.75,2) {};
            \node[lnode=${b_1}(u)$, green] (v1) [below left=0.85cm and 4.00cm of v0] {};
            \node[rnode=${b_2}(u)$, green] (v2) [below right=0.85cm and 4.00cm of v0] {};

            \node[lnode=${b_3}$, green] (v3) [below left=0.95cm and 1.60cm of v2] {};
            \remainingG{13.9,-0.8}{$G-N[v]-u$}
            \node[rnode=${b_4}$, green] (v4) [below right=0.95cm and 1.90cm of v2] {};
            \remainingG{17.9,-0.8}{$G-N[v,u]$}

            \node[lnode=${b_5}$, green] (v5) [below left=0.95cm and 1.90cm of v1] {};
            \remainingG{5.6,-0.8}{$G-\{v,u\}$}
            \node[rnode=${b_6}$, green] (v6) [below right=0.95cm and 1.60cm of v1] {};
            \remainingG{9.6,-0.8}{$G-v-N[u]$}

		      \draw[edge, -{Stealth[scale=1.5]}] (v0) to node[pos=0.36, above] {$v \notin I$} (v1);
		      \draw[edge, -{Stealth[scale=1.5]}] (v0) to node[pos=0.36, above] {$v \in I$} (v2);

		      \draw[edge, -{Stealth[scale=1.5]}] (v2) to node[pos=0.5, above left, inner sep=1pt] {$u \notin I$} (v3);
		      \draw[edge, -{Stealth[scale=1.5]}] (v2) to node[pos=0.5, above right, inner sep=1pt] {$u \in I$} (v4);

		      \draw[edge, -{Stealth[scale=1.5]}] (v1) to node[pos=0.5, above left, inner sep=1pt] {$u \notin I$} (v5);
		      \draw[edge, -{Stealth[scale=1.5]}] (v1) to node[pos=0.5, above right, inner sep=1pt] {$u \in I$} (v6);
        \end{scope}

	\end{tikzpicture}
    \end{adjustbox}
	\caption{This example illustrates the relationship between branching and the \enr{} framework for the given branching order $v$ followed by $u$. For each branch, an edge expansion occurs between the vertex from which we are branching and the vertex selected for branching. When we branch from the purely excluding branch (which has weight zero), the edge expansion generates a vertex that can be removed directly by domination. }\label{fig:branching}
\end{figure*}

While Theorem~\ref{thm:branching} bounds the number of new vertices introduced from branching, the fastest exact algorithms typically rely on reductions, applied either as preprocessing or interleaved with branching. For example, the $\mathcal{O}(2^{0.288n})$ algorithm by Fomin~\etal{}~\cite{fomin2006measure} employs reductions based on connected components, domination, and folding of low-degree vertices to obtain its running time. Similarly, practical solvers such as \textsc{KaMIS}~\cite{kamis2017} make extensive use of reduction rules and can handle real-world instances with millions of vertices.

The \enr{} framework also supports reductions within a branching-style process. Suppose a reduction rule determines that a vertex $a$ can be included in the solution for a branch represented by the vertex $b$. For that, \enr{} can perform an edge expansion between $a$ and $b$, introducing the vertex $v_{ab}$ with weight $\w(v_{ab}) = \w(a) + \w(b)$. Since $a$ can be included, the original vertex $b$ can be removed, and $v_{ab}$ becomes the representative of the branch, effectively extending the partial solution encoded in $b$ by the vertex $a$. Conversely, if a reduction rule determines that $a$ must be removed, the same expansion step is applied, but the newly created vertex $v_{ab}$ is immediately discarded. In this case, the vertex $b$ remains and becomes adjacent to $a$, ensuring that $a$ is removed from all extensions of the corresponding branch $b$.

\textbf{One-Lookahead Pruning.} Since \enr{} encodes branches as vertices, pruning a branch is the same as pruning a vertex that cannot belong to an optimal solution. Formally, let $\mathrm{LB}$ be the weight of the best-known independent set and $\textsc{UB}(G)$ be an upper bound on an MWIS in $G$. For any vertex $v$, if $\mathrm{LB} \geq \w(v) + \textsc{UB}(G - N[v])$, then $v$ can be safely removed. In Algorithm~\ref{alg:bnb2}, this pruning is used to discard branches, but in principle, any vertex in the graph can be reduced following this procedure.

The effectiveness of one-lookahead pruning heavily depends on the quality of the lower bound, as well as the tightness of the computed upper bound. For the lower bound, fast metaheuristics, such as iterated local search, can be used to obtain a high-quality solution early. For the upper bounds, clique covers are typically used in practice as an upper bound, since at most one vertex from each clique can be included in the solution~\cite{kamis2017}.

\begin{claim}\label{claim:enr_runtime}
    Any branching algorithm $A$ using only branching, reduction rules, and pruning compatible with \enr{} to solve MWIS in $\mathcal{O}^*(c^n)$ time can be adapted similarly to Algorithm \ref{alg:bnb1} with \hbox{running time $\mathcal{O}^*(c^n)$.}
\end{claim}

Let $n$ be the number of vertices in the original input graph. By Theorem~\ref{thm:branching}, simulating Algorithm $A$ creates at most $T = \mathcal{O}(c^n)$ branching vertices. At any point, the expanded graph has at most $n + T$ vertices. To ensure a running time of $\mathcal{O}^*(c^n)$, we must show that operations on any single subproblem take time polynomial in $n$, independent of $T$. 
We achieve this by having each original vertex maintain two distinct adjacency lists: one for its neighbors in the original graph and one for the branching states. Furthermore, the branching vertices are stored in a list according to Remark~\ref{rem:branching-clique} without explicitly enumerating the edges in that clique. This allows us to compute $G - N[b]$ in polynomial time in $n$. Consequently, any reduction rule, pruning step, or expansion applied to a subproblem operates on a graph of size at most $n$, and we require at most $\mathcal{O}(\text{poly}(n))$ time per node of the search tree. The total running time is therefore $\mathcal{O}(\text{poly}(n) \cdot c^n) = \mathcal{O}^*(c^n)$.

\section{Experimental Results.}


\begin{table*}[t]
\caption{Experimental results on a collection of instances that could not be completely reduced using the state-of-the-art method, learn-and-reduce. The last column $\alpha(G)$ gives the optimal solutions for the 10 new instances that \enr{} were able to reduce fully. The smallest $n_k$ and $m_k$ for each instance appear in bold. The highlighted rows indicate instances that Gurobi could solve optimally with the same time limit.}\label{tab:results}
\small
\setlength{\tabcolsep}{3.0pt}
\begin{tabular}{lrrrrrrrrrrr}
                         &          &              && \multicolumn{3}{c}{learn-and-reduce} && \multicolumn{3}{c}{expand-and-reduce} &                      \\ \cline{5-7} \cline{9-12}
\textbf{Instance}        & $n$      & $m$          && $n_k$    & $m_k$        & $t_{red}$ && $n_k$    & $m_k$        & $t_{red}$ & $\alpha(G)$             \\ \cline{1-3}\cline{5-7} \cline{9-12}
\addlinespace
\rowcolor{gray!15}
fe\,body  & 45\,087  & 163\,734     && 395      & 3\,551       & 0.6      && \textbf{0}        & \textbf{0}            & 317.9    & 1\,680\,182          \\
\rowcolor{gray!15}
fe\,pwt                  & 36\,519  & 144\,794     && 15\,239  & 106\,943     & 2.2      && \textbf{8\,868}   & \textbf{39\,232}      & 2\,804.3 &                      \\
fe\,rotor                & 99\,617  & 662\,431     && 89\,294  & 690\,242     & 5.1      && \textbf{88\,377}  & \textbf{568\,194}     & 3\,616.2 &                      \\
\rowcolor{gray!15}
osm\,alabama-3           & 3\,504   & 309\,664     && 815      & 61\,637      & 1.5      && \textbf{0}        & \textbf{0}            & 7.4      & 185\,744             \\
\rowcolor{gray!15}
osm\,california-3        & 587      & 27\,536      && 359      & 19\,705      & 0.2      && \textbf{327} & \textbf{16\,290} & 3\,615.9 &                      \\
\rowcolor{gray!15}
osm\,d.-of-c.-2          & 13\,597  & 1\,609\,795  && 2\,361   & 274\,529     & 5.8      && \textbf{1\,142}   & \textbf{150\,126}     & 554.5    &                      \\
\rowcolor{gray!15}
osm\,d.-of-c.-3          & 46\,221  & 27\,729\,137 && 26\,728  & \textbf{13\,406\,329} & 290.4    && \textbf{23\,687}  & 13\,631\,400 & 3\,607.7 &                      \\
\rowcolor{gray!15}
osm\,florida-3           & 2\,985   & 154\,043     && 658      & 41\,043      & 0.3      && \textbf{0}        & \textbf{0}            & 0.7      & 237\,333             \\
\rowcolor{gray!15}
osm\,georgia-3           & 1\,680   & 74\,126      && 481      & 31\,388      & 0.4      && \textbf{0}        & \textbf{0}            & 60.5     & 222\,652             \\
\rowcolor{gray!15}
osm\,greenland-3         & 4\,986   & 3\,652\,361  && 3\,402   & 2\,141\,096  & 27.4     && \textbf{3\,115}   & \textbf{2\,064\,291}  & 780.0    &                      \\
osm\,hawaii-3            & 28\,006  & 49\,444\,921 && 22\,849  & 38\,081\,016 & 936.0    && \textbf{20\,939}  & \textbf{35\,858\,776} & 3\,619.4 &                      \\
\rowcolor{gray!15}
osm\,idaho-3             & 4\,064   & 3\,924\,080  && 2\,920   & 2\,413\,139  & 40.0     && \textbf{2\,834}   & \textbf{2\,261\,042}  & 230.9    &                      \\
\rowcolor{gray!15}
osm\,kansas-3            & 2\,732   & 806\,912     && 1\,209   & 292\,997     & 3.8      && \textbf{639}      & \textbf{76\,002}      & 3\,664.2 &                      \\
osm\,kentucky-3          & 19\,095  & 59\,533\,630 && 15\,943  & 53\,739\,580 & 1\,510.4 && \textbf{15\,545}  & \textbf{53\,451\,091} & 3\,653.7 &                      \\
\rowcolor{gray!15}
osm\,massachusetts-3     & 3\,703   & 551\,491     && 1\,626   & 340\,652     & 3.0      && \textbf{1\,557}   & \textbf{335\,966}     & 337.1    &                      \\
\rowcolor{gray!15}
osm\,mexico-3            & 1\,096   & 47\,131      && 447      & 20\,443      & 0.3      && \textbf{0}        & \textbf{0}            & 0.9      & 97\,663              \\
\rowcolor{gray!15}
osm\,montana-3           & 837      & 69\,293      && 328      & 39\,859      & 0.3      && \textbf{0}        & \textbf{0}            & 5.4      & 59\,822              \\
\rowcolor{gray!15}
osm\,new-york-3          & 837      & 88\,728      && 526      & 50\,572      & 0.4      && \textbf{401}      & \textbf{28\,533}      & 3\,708.7 &                      \\
\rowcolor{gray!15}
osm\,north-carolina-3    & 1\,557   & 236\,739     && 997      & 133\,106     & 0.9      && \textbf{0}        & \textbf{0}            & 14.1     & 49\,720              \\
\rowcolor{gray!15}
osm\,oregon-3            & 5\,588   & 2\,912\,701  && 3\,159   & 1\,817\,045  & 23.8     && \textbf{2\,934}   & \textbf{1\,774\,335}  & 392.8    &                      \\
\rowcolor{gray!15}
osm\,rhode-island-2      & 2\,866   & 295\,488     && 498      & 46\,183      & 0.5      && \textbf{0}        & \textbf{0}            & 0.2      & 184\,596             \\
\rowcolor{gray!15}
osm\,rhode-island-3      & 15\,124  & 12\,622\,219 && 12\,189  & 11\,225\,470 & 153.1    && \textbf{11\,787}  & \textbf{11\,113\,933} & 2\,842.8 &                      \\
\rowcolor{gray!15}
osm\,vermont-3           & 3\,436   & 1\,136\,164  && 1\,946   & 526\,463     & 5.5      && \textbf{1\,708}   & \textbf{409\,654}     & 243.8    &                      \\
\rowcolor{gray!15}
osm\,virginia-3          & 6\,185   & 665\,903     && 2\,385   & 293\,068     & 2.5      && \textbf{1\,746}   & \textbf{231\,977}     & 482.1    &                      \\
\rowcolor{gray!15}
osm\,washington-3        & 10\,022  & 2\,346\,213  && 6\,671   & 1\,955\,568  & 14.6     && \textbf{5\,819}   & \textbf{1\,766\,225}  & 1\,186.5 &                      \\
\rowcolor{gray!15}
osm\,west-virginia-3     & 1\,185   & 125\,620     && 864      & 98\,399      & 0.5      && \textbf{0}        & \textbf{0}            & 2\,322.9 & 47\,927              \\
\rowcolor{gray!15}
snap\,as-skitter         & 1\,696\,415 &11\,095\,298 && \textbf{1\,951} & \textbf{27\,277}      & 3.4      && 5\,428   & 41\,676      & 3\,522.4 &                      \\
\rowcolor{gray!15}
snap\,loc-gowalla\,edges & 196\,591 &   950\,327  && \textbf{304}      & 2\,938       & 0.3      && 486   &   \textbf{2\,653}      & 3\,786.5    &                      \\
\rowcolor{gray!15}
snap\,soc-LiveJ.         & 4\,847\,571 & 42\,851\,237 && 1\,715   & 28\,390      & 23.5     &&  \textbf{1\,372} & \textbf{11\,182}  & 5\,302.4   &                      \\
snap\,soc-p.-rel         &1\,632\,803& 22\,301\,964 && \textbf{782\,676} & 14\,234\,808 & 644.0    && 963\,686 & \textbf{11\,160\,706}  & 3\,805.2 &          \\
\rowcolor{gray!15}
snap\,web-NotreDame      & 325\,729 & 1\,090\,108  && 97       & 1\,357       & 0.6      && \textbf{0}        & \textbf{0}            & 299.1    & 26\,016\,941         \\
\rowcolor{gray!15}
ssmc\,fl2010             & 484\,481 & 1\,173\,147  && 2\,045   & 8\,459       & 3.6      && \textbf{1\,668}   & \textbf{5\,768}       & 138.7    &                      \\
\end{tabular}
\end{table*}

In the following, we briefly describe our reference implementation for the reduction aspect of \enr{}. While the reduction rules domination and clique fold can be applied exhaustively, our framework does not enforce a specific order for edge expansions. Instead, this order depends on the specific algorithm being implemented. 

\textbf{Implementation.} We implement a simple unguided search for edge expansions that yield a smaller graph after applying domination and clique fold. Following Gellner et al.~\cite{gellner2021boosting}, we restrict edge expansions to a vertex's neighborhood. This localizes the graph blow-up and makes it easier to track candidate vertices for domination. If expanding and reducing yield a smaller graph, we continue from this smaller instance. Otherwise, we revert the changes and start from another vertex. See Algorithm~\ref{alg:framework} in the Appendix for details and hyperparameters. Our code is publicly available on GitHub\footnote{\url{https://github.com/KennethLangedal/ExpandAndReduce}} and archived on Zenodo\footnote{\url{https://doi.org/10.5281/zenodo.22960898}}.

\textbf{Experimental Setup.} The experiments were executed on a machine with an AMD Ryzen 9 5950X 16-Core processor, 32GB RAM, running Ubuntu 24.04 with Linux kernel 6.17.0-20. The code was written in C and compiled with GCC version 13.3 using the O3 optimization flag. We compared \enr{} against the current state-of-the-art baseline, learn-and-reduce~\cite{grossmann2025accelerating}, which incorporates a wide range of reduction rules and screening techniques to achieve faster running times. We use the reduced number of vertices and edges ($n_k$ and $m_k$) in addition to execution time as the primary evaluation metric. We use the same dataset used by learn-and-reduce, which includes large social networks from the Stanford Large Network Dataset Repository (snap)~\cite{snapnets}, real-world graphs from OpenStreetMaps (osm)~\cite{barth-2016,cai-dynwvc,OSMWEB}, instances from the SuiteSparse Matrix Collection (ssmc) \cite{davis2011university,ssmcWEB}, triangle meshes (mesh)~\cite{sander2008efficient}, and 3D meshes derived from simulations using the finite element method (fe)~\cite{walshaw}. To highlight the new reductions that \enr{} can achieve, we focus on a subset of these instances removing any graph that learn-and-reduce already reduces fully.

\textbf{Experimental Results.} The results of our experiments are shown in Table~\ref{tab:results}. Since \enr{} searches blindly for vertices to reduce, it is no surprise that \enr{} is much slower than learn-and-reduce. However, \enr{} is also significantly more effective at reducing the graph. Out of $32$ graphs previously not fully reduced, \enr{} fully reduces $10$, yielding $10$ new optimal solutions. Furthermore, in all but three instances, \enr{} produces smaller irreducible graphs than the baseline. This result experimentally verifies Theorem~\ref{thm:reductions_cover}, demonstrating that edge expansion, domination, and clique fold can simplify existing reduction rules. Additionally, it shows that there are still reducible patterns not addressed by the current rules. Our approach is also substantially simpler to implement. While exact code complexity is hard to quantify, a line count via \texttt{cloc}~\cite{adanial_cloc} shows that the baseline has over $20$ times more lines of code than \enr{}.

Finally, we evaluated the impact of \enr{} when combined with Gurobi~\cite{gurobi}, the industry-standard optimization solver. Interestingly, both learn-and-reduce and \enr{} had no impact on the number of instances solved by Gurobi. On instances where Gurobi reached the timeout threshold, it did so regardless of data reductions used. The four instances where Gurobi failed to solve within the one-hour time limit were also among the largest graphs, and they remained large even after reductions; see Table~\ref{tab:results}.

\section{Conclusion and Future Work.}

The biggest contribution of this paper is the new \enr{} framework. We demonstrate the effectiveness of \enr{} for MWIS, both theoretically and with preliminary experimental results. We also believe \enr{} to be a valuable contribution to other problems, assuming there are expansion rules for them. The edge expansion rule for MWIS is surprisingly simple and elegant, which may not be the case for other problems. One reason it works so well for MWIS is that we create cliques, and only one vertex from each clique can be in the solution. For most other problems, we would not have this direct connection. Therefore, a natural way to design expansion rules for other problems is to change the original problem to be given a graph and a partition of its vertices, and ask for a solution that uses at most one vertex from each part. 

The fact that we cover a wide variety of existing reductions for MWIS with only three rules makes it possible to simplify existing algorithms, both theoretical analysis and practical implementations. 
We also show that we can reduce graphs where previous reduction rules do not apply; further investigation in this direction is needed.
More importantly our results highlight new understanding of how branching and dynamic programming relates and provide an elegant way of unifying these techniques.

Just as with branching, where there is no set procedure for choosing the next branching vertex, we do not have a set procedure for choosing the next edge expansion. We have shown that branching algorithms and dynamic programming algorithms provide upper bounds for the number of edge expansions needed to solve MWIS. This demonstrates the potential of \enr{}. Our results imply that we can eliminate both branching and dynamic programming if we manage to find a good enough selection procedure for edge expansions.
As the exponential explosion comes from the number of edge expansions, improved bounds on this number could yield better algorithms. Therefore, we pose the following question: Which order of reductions and edge expansions can reduce the number of edge expansions needed to solve MWIS?

We believe the \enr{} framework holds great potential for experimental algorithms as edge expansions can be combined with many existing techniques. It promises the same results as before with a much simpler rule set. As an alternative to branching and dynamic programming, \enr{} enables reductions that were not previously possible, such as domination between branching states and reductions of partial solutions during dynamic programming.
We cannot easily minimize the number of edge expansions, but we can rely on a heuristic selection procedure guided by theoretical results.
We have also demonstrated experimentally that our three rules can discover new reductions by simply searching for sequences of the three rules that lead to a reduced graph. Our preliminary results show that \enr{} can outperform the state-of-the-art preprocessing method in terms of reduced graph size, yielding 10 new optimal solutions on instances commonly used in the literature.

These results lay the theoretical foundation for a new type of algorithm that combines the power of branching and dynamic programming in a simple form amenable to experimental implementation.
Hopefully, these results are of interest to multiple communities and can bring theory and practice one step closer.

\section*{Acknowledgments.}
We thank Saket Saurabh for his feedback and help with editing.

\section*{Disclosure of Generative AI Use.}
Generative AI tools were used to assist with language editing and stylistic refinement of the
manuscript. All scientific contributions were performed by the authors.

\bibliographystyle{siamplain}
\bibliography{biblio}

\newpage
\clearpage

\appendix

\section{Finishing Proof of Theorem~\ref{thm:reductions_cover}.}
\label{sec_app_finishing_proof}
In the following, we analyze the remaining rules from Table~\ref{tab:reduction_overview} and show how we can reduce the same patterns equivalently by using edge expansion, domination and clique fold, finishing the proof \hbox{of Theorem~\ref{thm:reductions_cover}.}

\begin{regel}[Basic Single-Edge by Gu et al.~\cite{gu2021towards}]\label{regel:se}
    Let $u,v\in V$ be adjacent vertices with $\omega(N(u)\setminus N(v)) \leq \omega(u)$, then we can remove $v$.
\end{regel}
\begin{proof}
    We apply edge expansions to make $N(u)\setminus N(v)$ into a clique following Lemma~\ref{lem:clique_expansion}. Since $\omega(N(u)\setminus N(v)) \leq \omega(u)$ all vertices in that clique have less weight than $\omega(u)$. By construction, all newly added vertices in that clique are adjacent to all neighbors of $u$ and $v$. Then, we can apply domination, since the vertex $u$ dominates all of these newly added vertices. We can additionally remove the vertex $v$ by that rule, since after removing all new vertices, it holds that $N(u)\subseteq N(v)$.
\end{proof}

\begin{regel}[Extended Single-Edge by Gu et al.~\cite{gu2021towards}]\label{regel:ese}
    Let $u,v\in V$ be adjacent vertices with $\omega(v)\geq \omega(N(v)) - \omega(u)$, then we can remove $N(u)\cap N(v)$.
\end{regel}

\begin{proof}
    Let $N=N(v)\setminus N[u]$ be the set of vertices only adjacent to $v$ in the original instance. We apply edge expansion to make $N(u)\cap N(v)$ into a clique $C$.    
    Afterward, we apply edge expansion to pairs of vertices $c \in C$ and $n\in N$. Let $N'$ be the set of new vertices created by this step. Now, we apply edge expansion to pairs of vertices in the set $N'$ and $N$. The vertices created by this step are added to the set $N'$. We repeat this process until, for all $w\in N'$, it holds that $N(v) \subseteq N(w)$. Note that this is almost making $N \cup C$ into a clique, except that no vertices we create represent an independent set exclusively from $N$. Similar to Lemma~\ref{lem:clique_expansion}, we know all vertices in $C\cup N'$ have a weight less than $\omega(v)$. By construction, it holds for all vertices $w\in C \cup N'$ that $N(v) \subseteq N(w)$. Therefore, the vertices in $C\cup N'$ are all dominated by $v$, and can be removed. This results in the described pattern, since all newly added vertices, as well as the intersection $N(v)\cap N(u)\subseteq C$, are removed, but no new edges are added \hbox{between remaining vertices.}
\end{proof}

The next reduction we analyze is the extended twin presented by Gro{\ss}mann~\etal~\cite{grossmann2025accelerating}. It generalizes the twin reduction introduced by Lamm~\etal~\cite{lamm2019exactly}. Two vertices $u,v \in V$ are called twins whenever their neighborhoods are equal. Twins are true whenever they are adjacent and otherwise false. Depending on the vertices' weights, false twins can be included or folded as described in the following rule.
\textbf{}

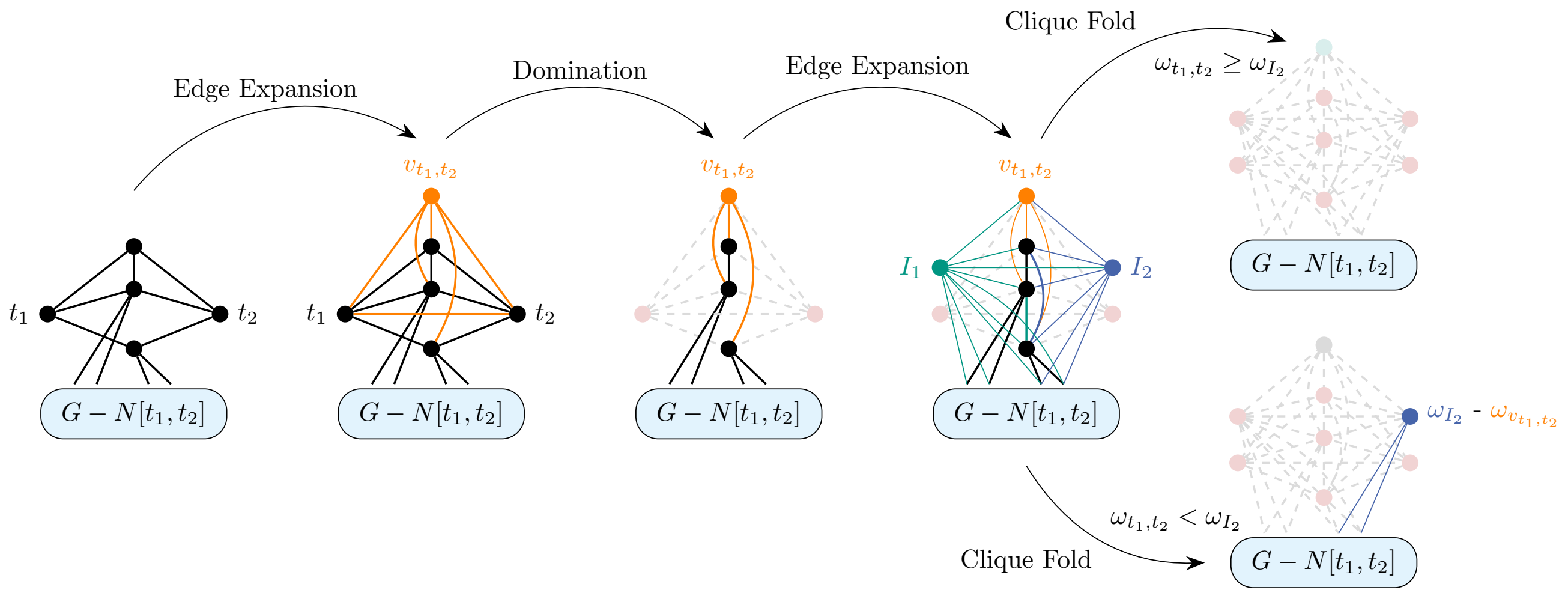
\begin{figure*}
	\centering
    \begin{adjustbox}{width=\textwidth}
	\begin{tikzpicture}
		\node[node] (v1) at (5,4.25) {};
		\node[lnode=$t_1$] (t1) [below left=.75cm and 1cm of v1] {};
		\node[rnode=$t_2$] (t2) [below right=.75cm and 1cm of v1] {};
    	\node[node] (v2) [below =0.35cm of v1] {};
    	\node[node] (v3) [below =1.15cm of v1] {};
		\remainingG{5,2}{$G-N[t_1, t_2]$}
		\draw[edge] (t1) -- (v1);
		\draw[edge] (t2) -- (v1);
		\draw[edge] (t1) -- (v2);
		\draw[edge] (t2) -- (v2);
		\draw[edge] (t1) -- (v3);
		\draw[edge] (t2) -- (v3);
		\draw[edge] (v2) -- (v1);
        \draw[edge] (v2) -- (4.5,2.4);
        \draw[edge] (v2) -- (4.2,2.4);
        \draw[edge] (v3) -- (5.5,2.4);
        \draw[edge] (v3) -- (5.2,2.4);
        \draw[bend left=40, -{Stealth[scale=1.5]}] (5,5) to  node[midway, above] {\nameref{edge_expansion}} (8.8,5.7);
        
	    \begin{scope}[xshift=4cm]
		\node[node] (v1) at (5,4.25) {};
		\node[lnode=$t_1$] (t1) [below left=.75cm and 1cm of v1] {};
		\node[rnode=$t_2$] (t2) [below right=.75cm and 1cm of v1] {};
    	\node[node] (v2) [below =0.35cm of v1] {};
    	\node[node] (v3) [below =1.15cm of v1] {};
    	\node[tnode={\textcolor{orange}{$v_{t_1,t_2}$}},orange] (t) [above =.45cm of v1] {};
		\remainingG{5,2}{$G-N[t_1,t_2]$}
		\draw[edge] (t1) -- (v1);
		\draw[edge] (t2) -- (v1);
		\draw[edge] (t1) -- (v2);
		\draw[edge] (t2) -- (v2);
		\draw[edge] (t1) -- (v3);
		\draw[edge] (t2) -- (v3);
		\draw[edge] (v2) -- (v1);
        \draw[edge] (v2) -- (4.5,2.4);
        \draw[edge] (v2) -- (4.2,2.4);
        \draw[edge] (v3) -- (5.5,2.4);
        \draw[edge] (v3) -- (5.2,2.4);
        \draw[edge,orange] (t1) -- (t2);
        \draw[edge,orange] (t) -- (t1);
        \draw[edge,orange] (t) -- (t2);
        \draw[edge,orange] (t) -- (v1);
        \draw[edge,orange,bend right] (t) to (v2);
        \draw[edge,orange,bend left] (t) to (v3);
        \draw[bend left=40, -{Stealth[scale=1.5]}] (5.2,5.7) to  node[midway, above] {\nameref{domination}} (8.8,5.7);
	    \end{scope}
        
	    \begin{scope}[xshift=8cm]        
		\node[node] (v1) at (5,4.25) {};
		\node[nodeE] (t1) [below left=.75cm and 1cm of v1] {};
		\node[nodeE] (t2) [below right=.75cm and 1cm of v1] {};
    	\node[node] (v2) [below =0.35cm of v1] {};
    	\node[node] (v3) [below =1.15cm of v1] {};
    	\node[tnode={\textcolor{orange}{$v_{t_1,t_2}$}},orange] (t) [above =.45cm of v1] {};
		\remainingG{5,2}{$G-N[t_1,t_2]$}
		\draw[edgeR] (t1) -- (v1);
		\draw[edgeR] (t2) -- (v1);
		\draw[edgeR] (t1) -- (v2);
		\draw[edgeR] (t2) -- (v2);
		\draw[edgeR] (t1) -- (v3);
		\draw[edgeR] (t2) -- (v3);
		\draw[edge] (v2) -- (v1);
        \draw[edge] (v2) -- (4.5,2.4);
        \draw[edge] (v2) -- (4.2,2.4);
        \draw[edge] (v3) -- (5.5,2.4);
        \draw[edge] (v3) -- (5.2,2.4);
        \draw[edgeR] (t1) -- (t2);
        \draw[edgeR] (t) -- (t1);
        \draw[edgeR] (t) -- (t2);
        \draw[edge,orange] (t) -- (v1);
        \draw[edge,orange,bend right] (t) to (v2);
        \draw[edge,orange,bend left] (t) to (v3);
        \draw[bend left=40, -{Stealth[scale=1.5]}] (5.2,5.7) to  node[midway, above] {\nameref{edge_expansion}} (8.8,5.7);
	    \end{scope}
        
	    \begin{scope}[xshift=12cm]        
		\node[node] (v1) at (5,4.25) {};
		\node[nodeE] (t1) [below left=.75cm and 1cm of v1] {};
		\node[nodeE] (t2) [below right=.75cm and 1cm of v1] {};
    	\node[node] (v2) [below =0.35cm of v1] {};
    	\node[node] (v3) [below =1.15cm of v1] {};
    	\node[tnode={\textcolor{orange}{$v_{t_1,t_2}$}},orange] (t) [above =.45cm of v1] {};
    	\node[lnode={\textcolor{green}{$I_1$}},green] (e1) [above=0.4cm  of t1] {};
    	\node[rnode={\textcolor{blue}{$I_2$}},blue] (e2) [above=0.4cm of t2] {};
		\remainingG{5,2}{$G-N[t_1,t_2]$}
		\draw[edgeR] (t1) -- (v1);
		\draw[edgeR] (t2) -- (v1);
		\draw[edgeR] (t1) -- (v2);
		\draw[edgeR] (t2) -- (v2);
		\draw[edgeR] (t1) -- (v3);
		\draw[edgeR] (t2) -- (v3);
        \draw[edgeR] (t1) -- (t2);
        \draw[edgeR] (t) -- (t1);
        \draw[edgeR] (t) -- (t2);
		\draw[edge] (v2) -- (v1);
        \draw[edge] (v2) -- (4.5,2.4);
        \draw[edge] (v2) -- (4.2,2.4);
        \draw[edge] (v3) -- (5.5,2.4);
        \draw[edge] (v3) -- (5.2,2.4);
        \draw[orange] (t) -- (v1);
        \draw[orange,bend right] (t) to (v2);
        \draw[orange,bend left] (t) to (v3);
		\draw[green] (e1) -- (e2);
		\draw[green] (e1) -- (v1);
		\draw[green] (e1) -- (v2);
		\draw[green] (e1) -- (v3);
		\draw[green] (e1) -- (t);
		\draw[blue] (e2) -- (v1);
		\draw[blue] (e2) -- (v2);
		\draw[blue] (e2) -- (v3);
		\draw[blue] (e2) -- (t);
		\draw[edge,green] (v2) -- (v3);
        \draw[edge,blue,bend left] (v1) to (v3);
        \draw[green] (e1) -- (4.5,2.4);
        \draw[green] (e1) -- (4.2,2.4);
        \draw[green] (e1) -- (5.2,2.4);
        \draw[green,bend left=20] (e1) to (5.5,2.4);
        \draw[blue] (e2) -- (5.5,2.4);
        \draw[blue] (e2) -- (5.2,2.4);
        \draw[bend left=40, -{Stealth[scale=1.5]}] (5.2,5.7) to  node[midway, above left] {\nameref{clique_fold}} node[midway, below right] {$\w_{t_1,t_2}\geq \w_{I_2}$} (8.5,7.);
        \draw[bend right=30, -{Stealth[scale=1.5]}] (5,1.3) to  node[midway, below left] {\nameref{clique_fold}} node[midway, above right] {$\w_{t_1,t_2}< \w_{I_2}$} (7.4,0.);
	    \end{scope}
        
	    \begin{scope}[xshift=16cm, yshift=2cm]        
		\node[nodeE] (v1) at (5,4.25) {};
		\node[nodeE] (t1) [below left=.75cm and 1cm of v1] {};
		\node[nodeE] (t2) [below right=.75cm and 1cm of v1] {};
    	\node[nodeE] (v2) [below =0.35cm of v1] {};
    	\node[nodeE] (v3) [below =1.15cm of v1] {};
    	\node[nodeI] (t) [above =.45cm of v1] {};
    	\node[nodeE] (e1) [above=0.4cm  of t1] {};
    	\node[nodeE] (e2) [above=0.4cm of t2] {};
		\remainingG{5,2}{$G-N[t_1,t_2]$}
		\draw[edgeR] (t1) -- (v1);
		\draw[edgeR] (t2) -- (v1);
		\draw[edgeR] (t1) -- (v2);
		\draw[edgeR] (t2) -- (v2);
		\draw[edgeR] (t1) -- (v3);
		\draw[edgeR] (t2) -- (v3);
        \draw[edgeR] (t1) -- (t2);
        \draw[edgeR] (t) -- (t1);
        \draw[edgeR] (t) -- (t2);
		\draw[edgeR] (v2) -- (v1);
        \draw[edgeR] (v2) -- (4.5,2.4);
        \draw[edgeR] (v2) -- (4.2,2.4);
        \draw[edgeR] (v3) -- (5.5,2.4);
        \draw[edgeR] (v3) -- (5.2,2.4);
        \draw[edgeR] (t) -- (v1);
        \draw[edgeR,bend right] (t) to (v2);
        \draw[edgeR,bend left] (t) to (v3);
		\draw[edgeR] (e1) -- (e2);
		\draw[edgeR] (e1) -- (v1);
		\draw[edgeR] (e1) -- (v2);
		\draw[edgeR] (e1) -- (v3);
		\draw[edgeR] (e1) -- (t);
		\draw[edgeR] (e2) -- (v1);
		\draw[edgeR] (e2) -- (v2);
		\draw[edgeR] (e2) -- (v3);
		\draw[edgeR] (e2) -- (t);
		\draw[edgeR] (v2) -- (v3);
        \draw[edgeR,bend left] (v1) to (v3);
        \draw[edgeR] (e1) -- (4.5,2.4);
        \draw[edgeR] (e1) -- (4.2,2.4);
        \draw[edgeR] (e1) -- (5.2,2.4);
        \draw[edgeR,bend left=20] (e1) to (5.5,2.4);
        \draw[edgeR] (e2) -- (5.5,2.4);
        \draw[edgeR] (e2) -- (5.2,2.4);
	    \end{scope}
        
	    \begin{scope}[xshift=16cm, yshift=-2cm]  
		\node[nodeE] (v1) at (5,4.25) {};
		\node[nodeE] (t1) [below left=.75cm and 1cm of v1] {};
		\node[nodeE] (t2) [below right=.75cm and 1cm of v1] {};
    	\node[nodeE] (v2) [below =0.35cm of v1] {};
    	\node[nodeE] (v3) [below =1.15cm of v1] {};
    	\node[nodeR] (t) [above =.45cm of v1] {};
    	\node[nodeE] (e1) [above=0.4cm  of t1] {};
    	\node[rnode={\textcolor{blue}{$\w_{I_2}$} - \textcolor{orange}{$\w_{v_{t_1,t_2}}$}},blue] (e2) [above=0.4cm of t2] {};
		\remainingG{5,2}{$G-N[t_1,t_2]$}
		\draw[edgeR] (t1) -- (v1);
		\draw[edgeR] (t2) -- (v1);
		\draw[edgeR] (t1) -- (v2);
		\draw[edgeR] (t2) -- (v2);
		\draw[edgeR] (t1) -- (v3);
		\draw[edgeR] (t2) -- (v3);
        \draw[edgeR] (t1) -- (t2);
        \draw[edgeR] (t) -- (t1);
        \draw[edgeR] (t) -- (t2);
		\draw[edgeR] (v2) -- (v1);
        \draw[edgeR] (v2) -- (4.5,2.4);
        \draw[edgeR] (v2) -- (4.2,2.4);
        \draw[edgeR] (v3) -- (5.5,2.4);
        \draw[edgeR] (v3) -- (5.2,2.4);
        \draw[edgeR] (t) -- (v1);
        \draw[edgeR,bend right] (t) to (v2);
        \draw[edgeR,bend left] (t) to (v3);
		\draw[edgeR] (e1) -- (e2);
		\draw[edgeR] (e1) -- (v1);
		\draw[edgeR] (e1) -- (v2);
		\draw[edgeR] (e1) -- (v3);
		\draw[edgeR] (e1) -- (t);
		\draw[edgeR] (e2) -- (v1);
		\draw[edgeR] (e2) -- (v2);
		\draw[edgeR] (e2) -- (v3);
		\draw[edgeR] (e2) -- (t);
		\draw[edgeR] (v2) -- (v3);
        \draw[edgeR,bend left] (v1) to (v3);
        \draw[edgeR] (e1) -- (4.5,2.4);
        \draw[edgeR] (e1) -- (4.2,2.4);
        \draw[edgeR] (e1) -- (5.2,2.4);
        \draw[edgeR,bend left=20] (e1) to (5.5,2.4);
        \draw[blue] (e2) -- (5.5,2.4);
        \draw[blue] (e2) -- (5.2,2.4);
	    \end{scope}
	\end{tikzpicture}
    \end{adjustbox}
	\caption{Illustration for how the extended twin reduction (Rule~\ref{regel:e_twin}) can be applied using only edge expansion, domination, and clique fold. In this example, it holds that $\w_{I_1} \leq \w_{I_2}$.}\label{fig:e_twin}
\end{figure*}
\begin{regel}[Extended Twin]\label{regel:e_twin}
    Let $t_1, t_2\in V$ be false twins and $\I_{N}$ be an MWIS on $G[N(t_1)]$.
    \begin{itemize}
        \item[(a)] If $\w(t_1)+\w(t_2) \geq \w(\I_{N})$, then $t_1$ and $t_2$ are includable.
        \item[(b)] If $\w(t_1) + \w(t_2) < \w(\I_{N})$ but $\I_{N}$ is the only independent set in $G[N(t_1)]$ with this property, then fold $t_1$, $t_2$, and $N(t_1)$ into $v'$. $G' = G[(V \cup \{v'\})\setminus N[\{t_1,t_2\}]]$ with $N(v') = N(N[\{t_1,t_2\}])$ and $\w(v') = \w(\I_{N}) - \w(t_1) - \w(t_2)$. We adjust the offset to $\alpha(G) = \alpha(G') + \w(t_1) + \w(t_2)$. If $v' \in \I'$, then $\I = (\I' \setminus \{v'\})\cup \I_{N}$, else $\I = \I' \cup \{t_1,t_2\}$.
        \item[(c)] Otherwise, fold $t_1$ and $t_2$ into a new vertex $v'$ with $N(v') = N(\{t_1,t_2\})$ and $\w(v') = \w(t_1) + \w(t_2)$. The offset stays the same $\alpha(G) = \alpha(G')$. If $v' \in \I'$, then $\I = \I' \setminus \{v'\} \cup \{t_1,t_2\}$, else $\I = \I'$.
    \end{itemize}
\end{regel}

\begin{proof}
    We first apply edge expansion to $t_1$ and $t_2$. This results in a clique $\{t_1,t_2,v_{t_1t_2}\}$. The newly added vertex $v_{t_1t_2}$ has the same neighborhood as $t_1$ and $t_2$, and it has a higher weight. Therefore, it dominates both $t_1$ and $t_2$, and we can remove these vertices according to domination, resulting in an equivalent reduced instance as in case (c).    
    For cases (a) and (b), we expand $N(v_{t_1t_2})$ into a clique following Lemma~\ref{lem:clique_expansion} using edge expansion. In the resulting graph, the vertex $v_{t_1t_2}$ is simplicial.
    In case (a), when $\alpha(G[N(v_{t_1t_2})]) = \w(\I_{N}) \leq \w(t_1) + \w(t_2) = \w(v_{t_1t_2})$, we know by Lemma~\ref{lem:clique_expansion}, that $v_{t_1t_2}$ is also of highest weight and we include $v_{t_1t_2}$ by applying clique fold. This corresponds to including $t_1$ and $t_2$ in the original graph. 
    In case (b), $v_{t_1t_2}$ is simplicial, but not of highest weight. Applying domination removes all but one vertex in the neighborhood of $v_{t_1t_2}$, since there is only one independent set in $N(t_1)=N(v_{t_1t_2})$ of higher weight than $\w(v_{t_1t_2})$.
    This vertex is representing the independent set $\I_N$, let it be $v_{\I_N}$. 
    We now can apply clique fold on $v_{t_1t_2}$, which folds $v_{t_1t_2}$ and reduces the weight of $v_{\I_N}$ to $\w(\I_{N})-\w(v_{t_1t_2}) = \w(\I_{N})-\w(t_1) - \w(t_2)$. With $v'=v_{t_1t_2}$, we get the same reduced instance. An illustration of these steps is given in Figure~\ref{fig:e_twin}.
\end{proof}

The survey~\cite{grossmann2024comprehensivesurveydatareduction} presents different variants of the struction rule. We only discuss the Extended Weighted Struction, since it is one of the less technical versions. The other versions all follow by the same arguments.
The main idea of these transformation rules is to reduce the stability number $\alpha$ of a graph, rather than the number of vertices. This is achieved by finding all independent sets in the neighborhood of a vertex $v$. These are then added as new vertices if their weight is higher than $\w(v)$.

\begin{regel}[Extended Weighted Struction by Gellner~et al.~\cite{gellner2021boosting}]\label{regel:struction}
    Let $v\in V$ and $C$ the set of all independent sets $c$ in $G[N(v)]$ with $\omega(v) < \omega(c)$. Then, we can transform the graph $G$ into $G'$ by the following steps.
    \begin{itemize}
        \item remove $N[v]$
        \item for each $c\in C$
        \begin{itemize}
            \item add a vertex $v_c$ with $\omega(v_c) = \omega(c) - \omega(v)$
            \item for each $w\in N(c)\setminus N(v)$ add the edge $\{w,v_c\}$
            \item for each $c'\in C\setminus \{c\}$, add the edge $\{v_c,v_{c'}\}$ forming a clique
        \end{itemize}
    \end{itemize}
    After this transformation, it holds that $\alpha(G) = \alpha(G') + \omega(v)$ and if a solution on the transformed instance $I'$ contains a vertex $v_c$ for some $c\in C$, then the original solution $I$ is constructed by $I = (I' \cap V) \cup c$. Otherwise, it holds $I = I' \cup \{v\}$.
\end{regel}
\begin{proof}
    We start by applying edge expansion to make the neighborhood of $v$ into a clique following Lemma~\ref{lem:clique_expansion}. After that, all independent sets in $C$ have been added as vertices to the graph and are connected with the edges described in the transformation steps. The vertices corresponding to independent sets $\tilde{c}$ with $\omega(v) \geq \omega(\tilde{c})$ are dominated by $v$ and can be removed by applying domination. After this step, we can apply clique fold to the vertex $v$. This removes $v$ from the graph, increases the offset so that $\alpha(G) = \alpha(G') + \omega(v)$ and reduces the weight of the vertices $v_c$ to $\omega(c) - \omega(v)$. The graph is now transformed to exactly the same as described by the above reduction rule.
\end{proof}

\begin{regel}[Heavy Set by Xiao~et al.~\cite{xiao2021efficient}]\label{regel:heavy_set}
    Let $u$ and $v$ be non-adjacent vertices having at least one common neighbor $x$. If for every independent set $\tilde{I}$ in the induced subgraph $G[N(\{u,v\})]$, $\omega(N(\tilde{I})\cap \{u,v\}) \geq \omega(\tilde{I})$, then include $u$ and $v$.
\end{regel}
\begin{proof}
    We apply edge expansion first to the vertices $u$ and $v$. This results in adding a new vertex $v_{uv}$. Now we proceed applying edge expansion on non-adjacent vertices in $N(v_{uv})$ until $N(v_{uv})$ forms a clique. Using Lemma~\ref{lem:clique_expansion}, we know $v_{uv}$ is simplicial. If for every independent set $\tilde{I}$ in the induced subgraph $G[N(\{u,v\})]$ it holds $\omega(N(\tilde{I})\cap \{u,v\}) \geq \omega(\tilde{I})$ this means that the vertices $u$, $v$ and $v_{uv}$ dominate all other vertices in their neighborhood. Furthermore, $v_{uv}$ is now simplicial and of highest weight. Therefore, using domination followed by clique fold results in the same reduction pattern.
\end{proof}

\begin{regel}[One Vertex Cut by Xiao et al. \cite{xiao2021efficient}]\label{regel:cut}
    Let $v$ be an articulation point and $G^*$ a connected component in $G-v$. Let $I_1$ be the optimal solution on $G^*$ and $I_2$ be the optimal solution on $G^*-N(v)$. If $\omega(v) + \omega(I_2) \leq \omega(I_1)$, then remove $v$ and include $I_1$. Otherwise, fold $v$ and $G^*$ into a new vertex $v'$ with weight $\omega(v')=\omega(v) +\omega(I_2)-\omega(I_1)$.
\end{regel}
\begin{proof}
    We start by applying edge expansion to make $V(G^*)$ into a clique. In this clique, all vertices except for $v$ are simplicial, and there is one vertex representing $I_1$, and one vertex representing $I_2$. We can apply domination until only the vertices $v_{I_1}, v_{I_2}$, and $v$ remain. 
    Since $I_2 \cap N[v] = \emptyset$, we can perform edge expansion between $v$ and $v_{I_2}$, resulting in a new vertex $x$ of weight $\omega(x) = \omega(v) + \omega(I_2)$ and neighborhood $N(x) = N(v) \cup N(v_{I_2})$. Since $N(v_{I_1}) \subseteq N(x)$, in the case of $\omega(v) + \omega(I_2) \leq \omega(I_1)$  $v_{I_1}$ dominates $x, v,$ and $v_{I_2}$. In the end for both cases, $v_{I_1}$ is simplicial and we apply clique fold on $v_{I_1}$, resulting in \hbox{the described pattern.}
\end{proof}

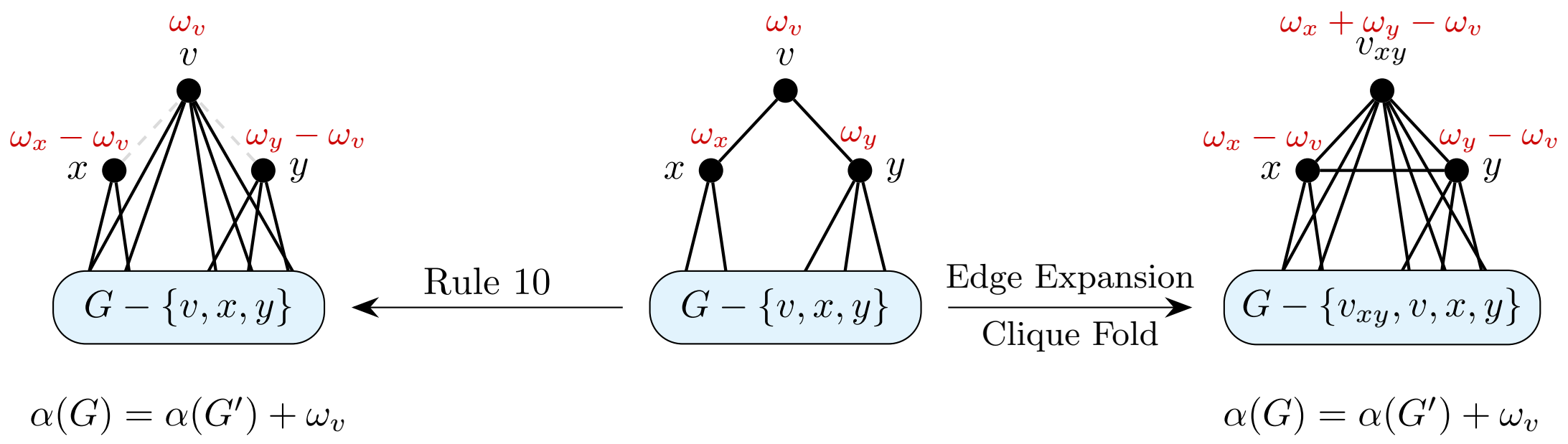
\begin{figure*}
    \centering
	\begin{tikzpicture}
        \draw[black,{Stealth[scale=1.5]}-] (6.5,2) -- (9,2) node[midway, above] {Rule \ref{regel:Vshape}};
        \draw[black,-{Stealth[scale=1.5]}] (12,2) -- (14.25,2) node[midway, above] {\small \nameref{edge_expansion}} node [midway, below] {\small \nameref{clique_fold}};
		\begin{scope}[xshift=5.5cm]
			\node[tnode=$v$] (v) at (5,4) {};
			\node[] (vtmp) [below=0.5cm of v] {};
			\node[rnode=$y$] (u) [right=0.45cm of vtmp]{};
			\node[lnode=$x$] (w) [left=0.45cm of vtmp]{};
    		\node[weight] (vweight) at ($(v) + (0, 0.3)$) {$\w_v$};
			\node[weight] (wweight) at (w) {$\w_x$};
			\node[weight] (uweight) at (u) {$\w_y$};
			\remainingG{5,2}{$G-\{v,x,y\}$}
			\draw[edge] (u) -- (6,2);
			\draw[edge] (u) -- (5.5,2);
			\draw[edge] (u) -- (5,2);
			\draw[edge] (w) -- (4.5,2);
			\draw[edge] (w) -- (4,2);
			\draw[edge] (u) -- (v);
			\draw[edge] (w) -- (v);
		\end{scope}
	    \node (offset) at (5,1) {$\alpha(G) = \alpha(G')+\w_v$};
	    \node[tnode=$v$] (v) at (5,4){};
		\node[] (vtmp) [below=0.5cm of v] {};
		\node[rnode=$y$] (u) [right=0.45cm of vtmp]{};
		\node[lnode=$x$] (w) [left=0.45cm of vtmp]{};
		\remainingG{5,2}{$G-\{v,x,y\}$}
		\draw[edge] (u) -- (6,2);
		\draw[edge] (u) -- (5.5,2);
		\draw[edge] (u) -- (5,2);
		\draw[edge] (w) -- (4.5,2);
		\draw[edge] (w) -- (4,2);
		\draw[edgeR] (u) -- (v);
		\draw[edgeR] (w) -- (v);
		\draw[edge] (v) -- (6.15,2);
		\draw[edge] (v) -- (5.7,2);
		\draw[edge] (v) -- (5.3,2);
		\draw[edge] (v) -- (4.3,2);
		\draw[edge] (v) -- (3.9,2);
		\node[weight] (vweight) at ($(v) + (0, 0.3)$) {$\w_v$};
		\node[weight] (wweight) at ($(w) + (-0.4, 0)$) {$\w_x - \w_v$};
		\node[weight] (uweight) at ($(u) + (0.4, 0)$) {$\w_y-\w_v$};
        
        \begin{scope}[xshift=11cm]
    	    \node (offset) at (5,1) {$\alpha(G) = \alpha(G')+\w_v$};
    	    \node[tnode=$v_{xy}$] (v) at (5,4){};
    		\node[] (vtmp) [below=0.5cm of v] {};
    		\node[rnode=$y$] (u) [right=0.45cm of vtmp]{};
    		\node[lnode=$x$] (w) [left=0.45cm of vtmp]{};
    		\node[weight] (vweight) at ($(v) + (.0, 0.3)$) {$\w_x + \w_y - \w_v$};
    		\node[weight] (wweight) at ($(w) + (-0.4, 0)$) {$\w_x - \w_v$};
    		\node[weight] (uweight) at ($(u) + (0.4, 0)$) {$\w_y-\w_v$};
    		\remainingG{5,2}{$G-\{v_{xy},v,x,y\}$}
    		\draw[edge] (u) -- (6,2);
    		\draw[edge] (u) -- (5.5,2);
    		\draw[edge] (u) -- (5,2);
    		\draw[edge] (w) -- (4.5,2);
    		\draw[edge] (w) -- (4,2);
    		\draw[edge] (u) -- (v);
    		\draw[edge] (w) -- (v);
    		\draw[edge] (w) -- (u);
    		\draw[edge] (v) -- (6.15,2);
    		\draw[edge] (v) -- (5.7,2);
    		\draw[edge] (v) -- (5.3,2);
    		\draw[edge] (v) -- (4.3,2);
    		\draw[edge] (v) -- (3.9,2);
        \end{scope}
	\end{tikzpicture}
    \caption{We illustrate equivalent instances after reducing the degree 2 v-shape pattern case $\omega(v) < \omega(x) \leq \omega(y)$. After applying Rule~\ref{regel:Vshape}, the resulting instance is shown on the left, while the equivalent instance achieved with our framework is shown on the right. There is a one-to-one mapping between the vertices in both instances, and the MWIS weight is the same in both, even though the weights and edges differ slightly.}
    \label{fig:vshape}
\end{figure*}

For some reductions, there is freedom in encoding different reduced patterns. This can be seen, for example, for the V-Shape rule introduced by Gu et al.~\cite{gu2021towards}. Figure~\ref{fig:vshape} shows two different states of how one can reduce the same pattern in different ways. Depending on subsequent solvers, one or the other might be preferable.
There are also reduction rules introduced in \cite{grossmann2025accelerating} that can add and remove edges, switching between similar states. One example is the following rule.

\begin{regel}[Extended Domination by Gro{\ss}mann~\etal~\cite{grossmann2025accelerating}]\label{regel:extended_dom}
    Let $u,v$ be adjacent vertices such that $N[u]\subseteq N[v]$. If $\omega(v) > \omega(u)$, then we can remove the edge between $u$ and $v$ if we update the weight of $v$ so that $\omega(v) = \omega(v) - \omega(u)$.
\end{regel}

We do not cover this rule in our framework and consider it more as a step between equivalent instances. Moreover, since our framework aims to build large cliques to find domination patterns, removing edges in this specific case is rather counterproductive. We do include this rule here only to show the equivalence between instances we arrive at after using the V-Shape rule.

\begin{regel}[V-Shape by Gu et al.~\cite{gu2021towards}]\label{regel:Vshape}
    Let $v\in V$ be a degree-two vertex with two non-adjacent neighbors $x,y\in V$ and $\omega(x) \leq \omega(y)$. If
    \begin{itemize}
        \item $\omega(v) \geq \omega(x)+\omega(y)$, include $v$
        \item $\omega(x) + \omega(y) > \omega(v) \geq \omega(y)$, then fold $v,x$, and $y$ into a new vertex $v'$ with weight $\omega(x) + \omega(y) - \omega(v)$
        \item $\omega(y) \geq \omega(v) \geq \omega(x)$, then fold $v$ into $y$ and add edges such that $N(x) = N(x) \cup N(y)$.  
        \item $\omega(x) > \omega(v)$, then we remove the edges $\{v,x\}$ and $\{v,y\}$, set the neighborhood of $v$ to $N(v) = N(x) \cup N(y)$, and $\omega(x) = \omega(x) - \omega(v)$, and $\omega(y) = \omega(y) - \omega(v)$.
    \end{itemize}
\end{regel}
\begin{proof} 
    We start by applying edge expansion to $x$ and $y$. That creates the new vertex $v_{xy}$ of weight $\omega(x) + \omega(y)$. Furthermore, $x$ and $y$ are now adjacent and $v_{xy}$ is also connected to $v$. Therefore, $v$ is simplicial. If
    \begin{itemize}
        \item $\omega(v) \geq \omega(x) + \omega(y)$: In this case, $x$, $y$, and $v_{xy}$ are dominated by $v$ and can removed. Then, we apply clique fold to include $v$.
        \item $\omega(x) + \omega(y) > \omega(v) \geq \omega(y)$: We apply domination to remove $x$ and $y$, then clique fold to $v$, which removes $v$ and sets the weight of $v_{xy}$ to $\omega(x) + \omega(y) - \omega(v)$.
        \item $\omega(y) \geq \omega(v) \geq \omega(x)$: $v$ dominates $x$ and can be removed. Next, we apply clique fold to $v$. The vertex $v_{xy}$ now has weight $\omega(x) + \omega(y) - \omega(v)$.
        Now we can use Rule~\ref{regel:extended_dom} applied to $y$ and the vertex $v_{xy}$. This is possible, since $\omega(y)<\omega(v_{xy})$ and $N(y)\subseteq N(v_{xy})$. This results in the new weight for $\omega(v_{xy}) = \omega(v_{xy}) - \omega(v) - (\omega(y) - \omega(v)) = \omega(v_{xy}) - \omega(y) = \omega(x)$ which is exactly the same instance described by the v-shape rule.
        \item $\omega(x) > \omega(v)$: We apply clique fold to $v$. This resulting pattern is shown in Figure~\ref{fig:vshape}. Both of these reduced instances are considered equivalent since they only use different ways of encoding the solutions.
    \end{itemize}%
\end{proof}

The following rule can not be covered equivalently by our framework. However, our framework can still conceptually perform the same reduction. While the unconfined rule removes a single vertex $v$, our framework would instead remove all vertices that contain $v$ in their independent sets. However, the independent sets needed to perform domination remain as vertices in our graph after the reduction.

\begin{regel}[Extended Unconfined by Gro{\ss}mann et al.~\cite{grossmann2025accelerating}]\label{regel:e_unconfined}
    For a set $S\subset V$, a vertex $x \in N(S)$ is a child of $S$, if $\omega(x)\geq \omega(S\cap N(x))$. If additionally $|N(x) \setminus N[S]| = 1$, we call it extending child. The vertex $y\in N(x)\setminus N[S]$ is the satellite from that extending child $x$.
    A vertex $v \in V$ can be removed if a contradiction can be obtained by the assumption that every MWIS contains $v$. This is done by the following procedure. We start with a set $S=\{v\}$ and search for a contradiction by repeatedly extending $S$ with satellites from an extending child until one of the following conditions hold. 
    \begin{itemize}
        \item There exists a child $x$ such that $\omega(x) \geq \omega(S\cap N(x)) + \alpha(G[N(x)\setminus N[S]])$.
        \item There exists no further satellites to extend $S$ with.
    \end{itemize}
    If this algorithms stops in the first case, $v$ is unconfined and can be removed.
\end{regel}
\begin{proof}
    For this rule, we first want to discuss the high-level idea before describing the detailed steps. This rule grows an independent set $S$ that is almost dominated by the corresponding extending children used to extend $S$. To cover this rule conceptually, we have to make sure that in the unconfined case, the vertex $v$, and all vertices representing independent sets that contain $v$ are removed. 

    Let $v_{S_{k+1}}$ be the vertex representing the current set $S$ with $|S|=k+1$, initially the vertex $v=v_{S_1}$. Furthermore, we enumerate the extending children by $x_i$ and the satellites by $s_i$ for $1\leq i \leq k$. 
    Whenever there exists a satellite $s_i$, we perform an edge expansion between $s_i$ and $v_{S_i}$ creating the vertex $v_{S_{i+1}}$. 

    For all $1\leq i \leq k$, a child $x$ of $S$ is either connected to the extending child $x_i$ and $v_{S_i}$ or not adjacent to $x_i$, since otherwise $|N(x_{i})\setminus N[S]| > 1$ and $x_i$ would not have been an extending child. Furthermore, after the last edge expansion to create $v_{S_{k+1}}$ it holds that $N(\{x_i | 1\leq i \leq k\})\subseteq N(v_{S_k})$. 

    Since for every $1 \leq i \leq k$ it holds that $\omega(x_i) \geq \omega(S_i\cap N(x_i))$, the weight $\omega(S_i \setminus N(x_i) \cup \{x_i\}) \geq \omega(S_i)$. By that, the vertex representing the set $S_i \setminus N(x_i) \cup \{x_i\}$ dominates $v_{S_i}$ after $v_{S_{i+1}}$ was created. We further want to show that we can construct an even better vertex dominating $v_{S_i}$, which is guaranteed not to contain $v$. 
    We now construct the independent set $X$ as follows. We start with the set $X=S_{k+1}$.
    Now, we trace back all paths from the last extending child via the corresponding satellites to the starting vertex $v$. Such a path always exists by construction. For all extending children $x_i$ along those paths from $x_k$ to $v$, we swap the vertices in $S_i\cap N(x_i)$ with the corresponding extending child $x_i$, i.e., $X = (X \setminus N(x_i)) \cup \{x_i\}$. This improves the weight of $X$ since $\omega(x_i) \geq \omega(S_i\cap N(x_i))$. This holds for all $1\leq i \leq k$. By this approach, we can, in the end, replace $v$ with an extending child of $v$.
    Furthermore, the set $X$ still contains all satellites that are not adjacent to later extending children.
    The vertex $v_X$ representing $X$ is created by the corresponding edge expansions. After $v_{S_{k+1}}$ was created, $v_X$ will dominate $v_{S_k}$ by construction.

    If there exists a child $x$ such that $\omega(x) \geq \omega(S\cap N(x)) + \alpha(G[N(x)\setminus N[S]])$, by this we can create the new vertex $v_X$ representing $X$ with $\omega(v_X) \geq \omega(S_k\setminus N(x) \cup \{x\}) \geq \omega(S_{k+1})$. After edge expansions between $v_{S_k}$ and all possible vertices in $N(x) \setminus N[S]$, it holds that $N(v_X) \subseteq N(\{x_i | 1\leq i \leq k\})\subseteq N(v_{S_{k+1}})$.

\end{proof}

\section{Experimental Setup and Results.}\label{sec:experimental_details}

This section provides additional details on our first reference implementation, the experimental setup, and the preliminary results. We stress that our reference implementation is only intended to demonstrate our claim that the three simple rules used in \enr{} can reduce as much as a much larger set of reduction rules. As evident by the execution times, randomly searching for good sequences of edge expansions is not as efficient as checking for the well-defined patterns described by existing reduction rules.

\begin{algorithm2e}
\SetCommentSty{itshape}
\DontPrintSemicolon
\caption{Unguided search for reductions in the expand-and-reduce framework. }
\label{alg:framework}

\SetKwFunction{FReduceLoop}{ReduceLoop}
\SetKwInOut{Input}{Input}

\Input{A graph $G = (V, E, \w)$, parameters $\Delta_s$ (start limit), $\Delta_i$ (increment), $L$ (small graph limit)}
\BlankLine

\For{$i \gets 0$ \KwTo $I_{\text{max}}$}{
    $\text{blowup\_limit} \gets \Delta_s + i \times \Delta_i$\;
    $\text{force} \gets |V| < L$\;
    
    \FReduceLoop{$G$, \text{blowup\_limit}, \text{force}}\;
    
    \tcp{Preserve completely removed vertices across restarts}
    $\texttt{RestartSearch}(G)$\;
}
\end{algorithm2e}

\begin{algorithm2e}
\SetCommentSty{itshape}
\DontPrintSemicolon
\caption{Supporting procedures for Algorithm~\ref{alg:framework}}
\label{alg:helpers}

\SetKwFunction{FExpandReduce}{ExpandAndReduce}
\SetKwFunction{FReduceLoop}{ReduceLoop}
\SetKwFunction{FEdgeExpand}{Expand}
\SetKwFunction{FReduce}{Reduce}

\SetKwProg{Fn}{Procedure}{:}{}
\Fn{\FReduceLoop{$G=(V,E,\w)$, $\text{limit}$, $\text{force}$}}{
    \For{$1$ \KwTo $N_{\text{inner}}$}{
        $P \gets \text{random permutation of } V$\;
        \ForEach{$u \in P$}{
            \FExpandReduce{$G, u, \text{limit}, \text{force}$}\;
        }
    }
}
\BlankLine

\SetKwProg{Fn}{Function}{:}{end}
\Fn{\FExpandReduce{$G=(V,E,\w), u, \text{limit}, \text{force}$}}{
    $n \gets |V|$\;
    $S \gets \texttt{SaveGraphState}(G)$\;
    
    \While{$|V| < n + \text{limit}$}{
        $v_1, v_2 \gets$ two distinct random non-adjacent neighbors of $u$\;
        $G \gets$ \FEdgeExpand{$G, v_1, v_2$}\;
        $G \gets$ \FReduce{$G$}\;
        
        \If{$|V| < n \textbf{ or } (\text{force} \textbf{ and } u \text{ is folded})$}{
            \Return $1$ \tcp*[r]{Success}
        }
    }
    $\texttt{RestoreGraphState}(G, S)$\;
    \Return $0$ \tcp*[r]{Fail}
}
\end{algorithm2e}

\textbf{Implementation.} Our reference implementation for the reduction component of \enr{} is detailed in Algorithm~\ref{alg:framework} and \ref{alg:helpers}. To enhance the readability of the core logic, we have extracted several hard-coded hyperparameters and replaced them with symbolic variables. Specifically, the outer loop in Algorithm~\ref{alg:framework} runs for a maximum of $I_{\text{max}} = 128$ iterations. The local blow-up limit begins at $\Delta_s = 4$ and increases by $\Delta_i = 4$ with each iteration. In Algorithm~\ref{alg:helpers}, the unguided reduction search is repeated $N_{\text{inner}} = 8$ times. To relax the blow-up limit for smaller graphs with sizes less than $L = 1024$, we utilize the \textit{force} flag, which keeps any reduction that allows $u$ to fold into its neighborhood. 

After the graph has been reduced, the resulting vertices represent independent sets in the original graph. We can project these independent sets back onto the unreduced input graph to identify vertices that have been completely removed. For instance, consider an edge expansion between vertices $a$ and $b$, which results in a new vertex $v_{ab}$. In this scenario, the vertex $a$ now exists in both the original vertex $a$ and the new vertex $v_{ab}$. For the original vertex $a$ to be completely removed, both $a$ and $v_{ab}$ must be removed. In the \textit{RestartSearch} function from Algorithm~\ref{alg:framework}, any vertex that has been completely removed in the reduced graph will also be removed after the restart. In other words, while any folded vertices or those representing an independent set are reset, any permanently removed vertices will remain removed across restarts.

\end{document}